\documentclass[aps,prd,onecolumn,showpacs,groupedaddress]{revtex4-2}
\usepackage{graphicx}
\usepackage{dcolumn}
\usepackage{bm}
\usepackage{color}
\usepackage{longtable}
\usepackage{supertabular}
\usepackage[T1]{fontenc}
\usepackage{epstopdf}
\usepackage{amsmath}
\usepackage{amssymb}
\usepackage{setspace}
\usepackage{multirow}
\usepackage{booktabs}
\usepackage[section]{placeins}
\usepackage{mathrsfs}
\usepackage{hyperref}

\makeatletter

\newcommand{\Rmnum}[1]{\expandafter\@slowromancap\romannumeral #1@} 
\newcommand{\bq}{\begin{equation}}
\newcommand{\eq}{\end{equation}}
\newcommand{\bqn}{\begin{eqnarray}}
\newcommand{\eqn}{\end{eqnarray}}
\newcommand{\nb}{\nonumber}

\makeatother

\begin{document}

\title{Nonflow suppression in flow analysis with a maximum likelihood estimator}

\author{Chong Ye$^{2, 1}$}
\author{Wei-Liang Qian$^{3, 4, 1}$}\email[E-mail: ]{wlqian@usp.br (corresponding author)}
\author{Cesar A. Bernardes$^{5}$}
\author{Sandra S. Padula$^{6}$}
\author{Rui-Hong Yue$^{1}$}\email[E-mail: ]{rhyue@yzu.edu.cn (corresponding author)}
\author{Yutao Xing$^{2}$}\email[E-mail: ]{xy@id.uff.br (corresponding author)}
\author{Takeshi Kodama$^{7, 2}$}

\affiliation{$^{1}$ Center for Gravitation and Cosmology, School of Physical Science and Technology, Yangzhou University, 225009, Yangzhou, China}
\affiliation{$^{2}$ Instituto de F\'isica, Universidade Federal Fluminense, 24210-346, Niter\'oi, RJ, Brazil}
\affiliation{$^{3}$ Escola de Engenharia de Lorena, Universidade de S\~ao Paulo, 12602-810, Lorena, SP, Brazil}
\affiliation{$^{4}$ Faculdade de Engenharia de Guaratinguet\'a, Universidade Estadual Paulista, 12516-410, Guaratinguet\'a, SP, Brazil}
\affiliation{$^{5}$ Instituto de F\'isica, Universidade Federal do Rio Grande do Sul, 91501-970, Porto Alegre, RS, Brazil}
\affiliation{$^{6}$ Instituto de F\'isica Te\'orica, Universidade Estadual Paulista, 01140-070, S\~ao Paulo, SP, Brazi}
\affiliation{$^{7}$ Instituto de F\'isica, Universidade Federal do Rio de Janeiro, 21945-970, Rio de Janeiro-RJ , Brazil}

\begin{abstract}
We show that the maximum likelihood estimator (MLE) is an effective tool for mitigating non-flow effects in flow analysis. 
To this end, one constructs two toy models that simulate non-flow contributions corresponding to particle decay and momentum conservation, respectively.
The performance of MLE is analyzed by comparing it against standard approaches such as particle correlation and event plane methods.
For both cases, MLE is observed to provide a reasonable estimate of the underlying flow harmonics, and in particular, its performance can be further improved when the specific form of the likelihood in the presence of non-flow can be assessed. 
The dependencies of extracted flow harmonics on the multiplicity of individual events and the total number of events are analyzed. 
Additionally, it is shown that the proposed approach performs efficiently in addressing deficiencies in detector acceptance.
These findings suggest MLE as a compelling alternative to standard methods for flow analysis.
\end{abstract}

\date{Dec. 29th, 2025}

\maketitle
\newpage

\section{Introduction}\label{section1}

Quark-gluon plasma (QGP)~\cite{qgp-review-12, qgp-review-13} is a phase of nuclear matter featured by the asymptotic freedom of quantum chromodynamics.
Such an extremely hot and dense medium can be formed through relativistic heavy-ion collisions~\cite{RHIC-star-overview-1, RHIC-brahms-overview-1, RHIC-phenix-overview-1, RHIC-phobos-overview-1, LHC-alice-review-01, LHC-atlas-review-01, LHC-cms-review-01}.
Phenomenologically, relativistic hydrodynamics is a widely adopted theoretical framework for modeling the evolution of QGP formed in such collisions~\cite{hydro-review-04, hydro-review-05, hydro-review-06, hydro-review-07, hydro-review-08, hydro-review-09, hydro-review-10}.
The approach treats the plasma as a continuum, capturing the underlying physics that explains observable phenomena.
Relevant observables include particle spectra at low and intermediate transverse momentum, flow harmonics, and particle correlations.
Measurements of azimuthal distributions have revealed the concept of a {\it perfect} liquid, first observed at RHIC~\cite{RHIC-star-v2-01}.
Azimuthal anisotropy has become a crucial observable for extracting system properties~\cite{RHIC-brahms-v2-01, RHIC-phenix-v2-01, RHIC-star-v2-05, LHC-alice-vn-01, LHC-atlas-vn-01, LHC-cms-vn-01}.
Further studies have extended these analyses to small systems~\cite{LHC-cms-vn-04, LHC-atlas-vn-07, LHC-small-system-review-02}, deformed nuclei~\cite{RHIC-star-v2-10, hydro-vn-deformed-07, hydro-vn-deformed-10, hydro-vn-deformed-11, hydro-vn-deformed-12} and recently radial flow fluctuations and correlations~\cite{LHC-atlas-vn-20}.
Hydrodynamic evolution primarily reflects the dynamic response to fluctuating initial conditions.
Its nonlinear nature has motivated extensive research on IC and final-state azimuthal anisotropies~\cite{hydro-v3-02,hydro-vn-33,sph-vn-03,hydro-vn-34,sph-vn-04,sph-vn-06,hydro-vn-45,sph-corr-30,sph-corr-33}.

Flow harmonics $v_n$ characterize the one-particle azimuthal anisotropic particle distributions in the momentum space~\cite{event-plane-method-1}: 
\begin{eqnarray}
f_1(\phi)=\frac{1}{2\pi}\left[1+\sum_{n=1}^{} 2v_{n}\cos{n(\phi-\Psi_n)}\right],
\label{oneParDis}
\end{eqnarray}
where $\phi$ is the azimuthal angle and $\Psi_n$ is the event plane for harmonic order $n$.
The elliptic flow ($v_2$) arises from almond-shaped overlap geometry~\cite{hydro-vn-08}, while triangular flow ($v_3$) is attributed to the initial state fluctuations~\cite{hydro-v3-01}.

Traditional methods like event plane techniques estimate $\Psi_n$ in order to evaluate the harmonic coefficients in Eq.~\eqref{oneParDis}~\cite{event-plane-method-1, event-plane-method-2}.
This approach relies on the reaction plane, which is not directly measured.
Other methods are primarily based on the particle correlations and the notions of the Q-vectors and cumulants~\cite{hydro-corr-ph-03, hydro-corr-ph-10, pythia-vn-10}.
One advantage of particle correlation is that the event planes in Eq.~\eqref{oneParDis} are canceled out in the resulting formalism.
Moreover, the cumulant can be expressed concisely in terms of the generating function.
This class of approaches includes the particle cumulant~\cite{hydro-corr-ph-03, hydro-corr-ph-04, hydro-corr-ph-10, hydro-corr-ph-23, hydro-corr-ph-27}, Lee-Yang zeros~\cite{hydro-corr-LY-zeros-01, hydro-corr-LY-zeros-02, hydro-corr-LY-zeros-03}, and symmetric cumulants~\cite{hydro-corr-ph-36}, among other developments~\cite{hydro-corr-ph-38, hydro-corr-ph-42, hydro-vn-pca-01}.
It is worth noting that the notation of multi-particle cumulants involves different definitions for flow harmonics associated with moments of different orders. 

Recently, it was proposed that the maximum likelihood estimator (MLE) provides an alternative for evaluating flow harmonics~\cite{sph-vn-10}.
The method treats harmonic coefficients as parameters of an underlying hypothetical probability distribution and employs the MLE, a standard {\it estimator} in statistical inference~\cite{book-statistical-inference-Wasserman}, to extract those parameters from the data.
Specifically, the estimated parameters, namely, the flow harmonics, ensure that the observed data are the most probable.
The proposed method has been employed to investigate the differential flow, flow factorization, and event-plane fluctuations~\cite{sph-vn-11}.
In particular, it has recently been suggested that the high-order and mixed flow factorization ratios are sensitive quantities to the initial state granularities regarding the MLE approach~\cite{sph-corr-33}.

A key characteristic of flow analysis is that each event generated in a realistic collision process contains only a finite number of particles.
This factor potentially introduces a sizable uncertainty of purely statistical nature when estimating the flow harmonics, even if hadrons are emitted independently according to a well-defined one-particle distribution function~\cite{sph-corr-32}.
Specifically, for events with lower multiplicity, multi-particle correlations are likely to be significantly affected by statistical uncertainty~\cite{sph-corr-31}.
This is intrinsically different from the uncertainty caused by initial condition fluctuations, which usually originate from the underlying microscopic description.
From a hydrodynamic perspective, the latter manifests as fluctuations in the initial energy density distribution on an event-by-event basis.
In practice, for a given estimator, statistical uncertainty often can be derived analytically~\cite{sph-vn-10}.
In this regard, MLE benefits from its inherited mathematical properties. 
To be specific, as an asymptotically normal and unbiased estimator, MLE's efficiency is guaranteed in the sense that it is more accurate than any other estimator at the limit of a significant sample size.
Besides, it is more flexible to introduce additional ansatzes in the flow distribution function, and therefore it is more potent in dealing with specifically tailored scenarios.
In terms of convergence, the method is either unbiased or asymptotically unbiased. 
The context of relativistic heavy-ion collisions meets most of MLE's demands, as measurements performed at RHIC and LHC have accumulated significant data for different collision systems at various centralities.

Another key aspect of this framework is the presence of non-flow effects~\cite{hydro-corr-non-flow-01, hydro-corr-non-flow-04, hydro-corr-ph-09, hydro-corr-non-flow-review-03}, in which the collective behavior cannot be fully described by independent particle emission through the one-particle distribution function. 
Phenomena such as resonance decay~\cite{hydro-corr-non-flow-04}, jet shower~\cite{qgp-review-15, qgp-review-20}, string fragmentation~\cite{jet-fragmentation-02}, Hanbury-Brown-Twiss interferometry~\cite{hbt-20}, back-to-back correlations~\cite{bbc-02}, and final-state interactions~\cite{hbt-20} all potentially contribute to non-flow. 
Moreover, conservation laws, such as global momentum conservation, also introduce deviations in particle correlations compared to scenarios governed purely by flow dynamics. 
By taking momentum conservation into account as a constraint, particle correlations can be approximately estimated using either the central limit theorem~\cite{hydro-corr-non-flow-01} or the saddle-point approximation~\cite{hydro-corr-non-flow-02}. 
Various efforts have been made along this line of research over the last decade~\cite{hydro-corr-non-flow-01, hydro-corr-non-flow-02, hydro-corr-non-flow-03, hydro-corr-non-flow-07, hydro-corr-non-flow-08, hydro-corr-non-flow-13, hydro-corr-non-flow-14, hydro-corr-non-flow-20}. 
While the impact of non-flow is probably negligible in collisions of heavy nuclei at LHC and top RHIC energies, explicit calculations~\cite{hydro-corr-non-flow-14, hydro-corr-non-flow-20} have shown that the effect may become sizable in smaller systems. 
Consequently, there has been renewed interest in non-flow analysis in recent years~\cite{hydro-corr-non-flow-review-03}.

Notably, the impact of non-flow is reduced in higher-order multiparticle cumulants~\cite{hydro-corr-ph-03, hydro-corr-ph-04}, a behavior attributed to the statistical suppression of uncorrelated signals.  
In the case of MLE, all emitted particles are involved in evaluating the likelihood, which typically scales more strongly than the corresponding cumulant order.  
In practice, MLE still assumes independent and identically distributed (i.i.d.) one-particle distributions that do not explicitly account for non-flow contributions.  
Nevertheless, due to the typically large event multiplicities, it is plausible that non-flow effects play only a minor role.  
This has motivated conjectures that MLE may inherently mitigate non-flow artifacts~\cite{sph-vn-10, sph-vn-10}.  
However, systematic studies of non-flow suppression within the MLE framework are still lacking in the literature.  
A thorough comparative analysis contrasting MLE with conventional approaches would thus help clarify its efficacy in isolating genuine collective phenomena.

The present study is motivated by the above considerations.
We conduct a comprehensive study to demonstrate that the MLE is an effective tool for suppressing non-flow effects in extracting harmonic coefficients.
We develop two toy models that simulate the non-flow contributions.
The performance of MLE is analyzed by comparing it against standard approaches such as particle correlation and event plane methods. 
For both cases, MLE is observed to provide a relatively better estimation.
In particular, MLE's performance can be further improved when the impact on the likelihood due to non-flow can be reasonably estimated.
The dependencies of extracted flow harmonics on the multiplicity of individual events and the total number of events are analyzed.
Additionally, it is shown that the proposed approach performs efficiently in addressing deficiencies in detector acceptance. 

The remainder of this paper is organized as follows.
In the next section, we provide a brief review of the MLE-based flow analysis scheme.
In Sec.~\ref{section3}, we introduce two toy models, aiming to mimic two non-flow effects related to particle decay and momentum conservation.
Numerical studies are carried out in Sec.~\ref{section4}, where we investigate the performance of the MLE approach.
The results are compared with those obtained using cumulants and the event-plane method.
We analyze the dependencies of our results on the multiplicity and event number, as well as in the context of detector deficiency. 
The final section provides further discussions and concluding remarks.

\section{Flow estimator based on maximum likelihood}\label{section2}

The most employed approaches for extracting flow harmonics from the data are based on multi-particle correlations~\cite{event-plane-method-2}.
They essentially rely on the following definition for $k$-particle correlator at the limit of infinite multiplicity~\cite{hydro-corr-ph-23}: 
\begin{eqnarray}
\langle k\rangle_{n_1,\cdots,n_k} \equiv \langle e^{i(n_1\phi_1 + \cdots + n_k\phi_k)} \rangle
= v_{n_1}\cdots v_{n_k} e^{i\left(n_1\Psi_{n_1}+\cdots+n_k\Psi_{n_k}\right)} , \label{nkCorr}
\end{eqnarray}
where $\langle\cdots\rangle$ denotes the average over distinct tuples of particles, assuming independent particle emission as described by Eq.~\eqref{oneParDis}.
If all the event planes involved in the correlator are identical, by choosing a specific set of indices $(n_1, \cdots, n_k)$ such that $\sum_{j=1}^k n_j = 0$, the product of flow harmonics $v_n$ is singled out by canceling out the event planes in the exponential.
This strategy furnishes a formalism independent of the event-plane angles~\cite{hydro-corr-ph-27}.
The simplest case is the two-particle correlation ($k=2$), where $n_1 = -n_2 = n$, for which it yields
\begin{eqnarray}
\langle 2 \rangle_{n,-n} \equiv \left \langle e^{in(\phi_{1}-\phi_{2})} \right \rangle
= \langle\cos{n(\phi_{1}-\phi_{2})} \rangle = v_n^2 ,
\label{eq2}
\end{eqnarray}
which directly relates the two-particle correlation to the square of the flow harmonic $v_n$.

As discussed above, for realistic scenarios, the multiplicity of a given event is finite.
The average in the second equality of Eq.~\ref{nkCorr} must be performed by enumerating all possible tuples constructed from the measured particles.
Instead of integration, the average is implemented by a summation that involves all possible combinations from $M$ discrete azimuthal angles.
Specifically, the integration employed in the third equality of Eq.~\eqref{eq2} is no longer viable.
For a finite number of particles $M$, it has to be replaced by the following summation: 
\begin{eqnarray}
\widehat{v_n^2} = \frac{1}{M(M-1)}\sum_{i\ne j} \cos n(\phi_i - \phi_j) .
\label{eqEst2}
\end{eqnarray}

In the context of statistical inference, Eq.~\eqref{eqEst2} is known as a statistical estimator, indicated by the circumflex.
As a function of the azimuthal angles, it is per se governed by a probability distribution effectively governed by Eq.~\eqref{oneParDis} that generates $\phi_i$s.
It can be shown that it is an unbiased estimator for $v_n^2$, but its square root furnishes a biased estimation for $v_n$.
The second moment, namely, the variance, of the distribution Eq.~\eqref{eqEst2} measures the uncertainty of the estimation, which is found to be~\cite{hydro-corr-ph-36}
\begin{eqnarray}
      \mathrm{Var}\left[\widehat{v_n^2}\right] = \frac{1+v_{2n}^2}{M(M-1)}+2\frac{M-2}{M(M-1)}v_{n}^{2}(1+v_{2n}) +\frac{(M-2)(M-3)}{M(M-1)}v_{n}^{4}-v_{n}^{4} ,\label{eqVar3}
\end{eqnarray}
which remains finite and decreases with increasing multiplicity, as expected.
Notably, the above notion generalizes to most recipes for deriving flow harmonics using particle correlators, in terms of $Q$-vectors~\cite{hydro-vn-07} or flow vectors~\cite{hydro-corr-ph-23}, which can be effectively viewed as estimators.
The variance of these quantities largely remains finite~\cite{sph-corr-31}, indicating an inherent statistical uncertainty due to the finite number of measured events and particle multiplicity.
A finite variance implies that uncertainties in the estimated flow harmonics are inevitably influenced by limited statistics, particularly for a finite number of events.
Therefore, such limitations of statistical origin must be considered when comparing results obtained from different flow measurement methods.

Following this line of thought, MLE was proposed as an alternative estimator for flow and related observables~\cite{sph-vn-10, sph-vn-11}.
For a given set of observations $y\equiv (y_1, y_2, \cdots, y_M)$, one assumes that they are sampled from an underlying joint probability distribution governed by several unknown parameters $\theta \equiv (\theta_1, \theta_2, \cdots, \theta_m)$.
The likelihood function $\mathcal{L}$ for the observed data is considered accessible and given by:
\begin{eqnarray}
\mathcal{L}(\theta) \equiv \mathcal{L}(\theta; y) = f(y; \theta) ,\label{defLn}
\end{eqnarray}
which is evaluated for the given observations as a function of the parameters $\theta$.
The intuition of MLE is to seek out the parameter sets for which the observed data attains the highest joint probability, namely:
\begin{eqnarray}
\widehat{\theta}_{\mathrm{MLE}}=\arg\max\limits_{\theta\in\Theta}\mathcal{L}(\theta) , \label{defMLE}
\end{eqnarray}
where $\Theta$ is the domain of the parameters.
In particular, for i.i.d. random variables, $f(y; \theta)$ is given by a product of likelihood functions $f^\mathrm{uni}$:
\begin{eqnarray}
f(y; \theta)=\prod_{j=1}^M f^\mathrm{uni}(y_j; \theta) . \label{iidfn1}
\end{eqnarray}

This scheme can be readily applied to collective flow analysis in heavy-ion collisions.
Considering an event consisting of $M$ particles, the likelihood function reads:
\begin{eqnarray}
\mathcal{L}(\theta; \phi_{1}, \cdots, \phi_M) = f(\phi_{1},\cdots, \phi_M; \theta)=\prod_{j=1}^{M}f_1(\phi_j; \theta) ,
\label{eqlikelihood}
\end{eqnarray}
where the one-particle distribution function is given by Eq.~\eqref{oneParDis}.
The parameters of the resulting joint distribution, $\theta = (v_1, v_2, \cdots, \Psi_1, \Psi_2, \cdots)$, are the flow harmonics and the event planes.

For computational efficiency, one often chooses the objective function to be the log-likelihood function ${\ell}$:
\begin{eqnarray}
{\ell}(\theta; \phi_{1}, \cdots, \phi_M) = \log\mathcal{L}(\theta; \phi_{1}, \cdots, \phi_M)
= \sum_{j=1}^{M}\log f_1(\phi_j; \theta) ,
\label{eqlogl}
\end{eqnarray}
as the maximum of $\ell$ coincides with that of $\mathcal{L}$.
For $\ell$ that is differentiable in its domain $\Theta$, the necessary conditions for the occurrence of a maximum are:
\begin{eqnarray}
\frac{\partial{\ell}}{\partial\theta_1}=\cdots=\frac{\partial{\ell}}{\partial\theta_m}=0 .
\label{condMLE}
\end{eqnarray}
The accuracy of MLE, the mean squared error, is governed by the Cramér-Rao lower bound for large sample size, which is superior to any other consistent estimator.
In the case of relativistic heavy-ion collisions, all events of a given large multiplicity constitute a multivariate normal distribution, whose standard deviation is expected to be asymptotically proportional to $\frac{1}{\sqrt{M}}$.
Nonetheless, for realistic events, the performance of MLE is still limited by the number of particles emitted from the collisions.

\section{Two toy models for non-flow effects}\label{section3}

To focus on the non-flow effects, we elaborate on two toy models tailored to simulate particle decay and momentum conservation.
The toy models are constructed in a minimal fashion, while the non-flow effects will be artificially exaggerated.
The calculations are limited to a Bjorken invariant scenario, so that only the azimuthal angles of the particles are considered. 
The goal is to conduct a comparative analysis that contrasts MLE with conventional approaches.

\begin{figure}[ht]
    \centering
    \begin{minipage}{0.4\textwidth}
        \centering
        \includegraphics[width=1.1\textwidth, height=0.26\textheight]{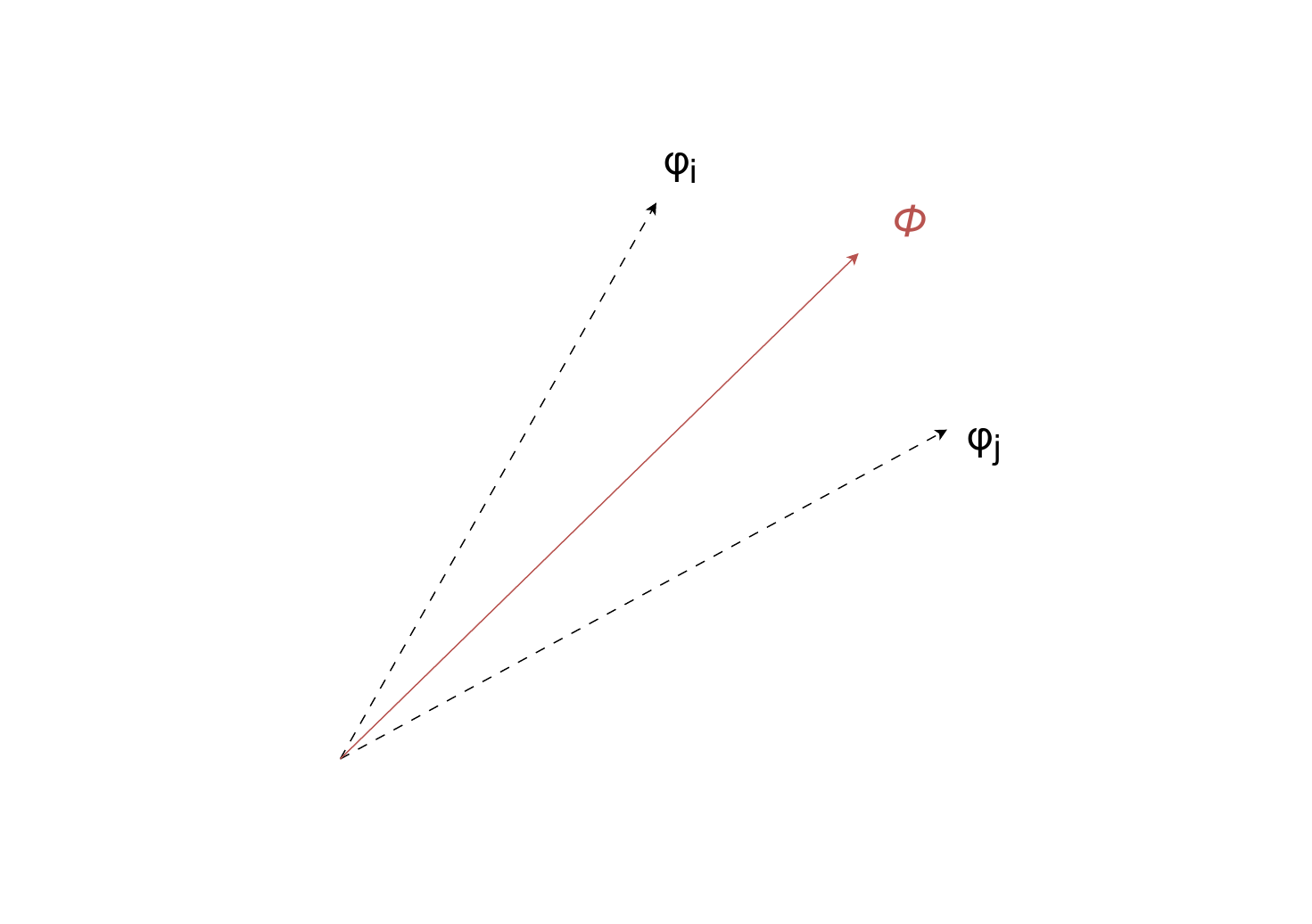}
    \end{minipage}
    \begin{minipage}{0.5\textwidth}
        \centering
        \includegraphics[width=1.0\textwidth, height=0.3\textheight]{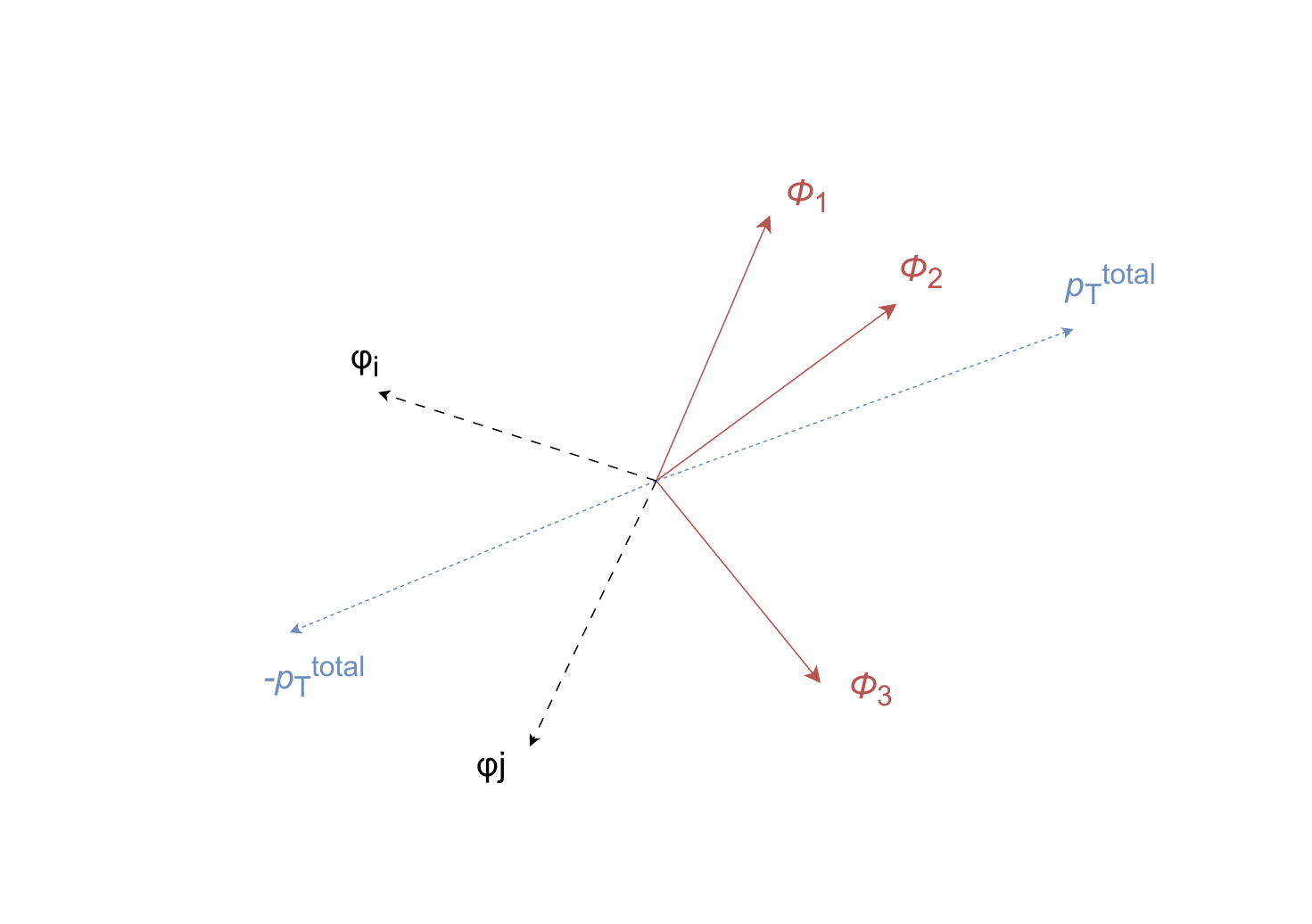}
    \end{minipage}
\renewcommand{\figurename}{Fig.}   
\caption{The two toy models devised in the present study.
Left: The first scenario mimics particle decays. 
Each particle might randomly split into two daughter particles separated by an angle drawn from a Gaussian distribution Eq.~\eqref{GDisDecay}.
Right: The second scenario incorporates momentum conservation. 
The emission of additional particles guarantees the total momentum conservation.
To enhance the non-flow effect, the conservation is enforced for all individual subsets consisting of only a few particles.}
\label{fig_toymodel_scheme}
\end{figure}

The first toy model considers the effect of particle decay.
One first generates an event consisting of a collection of first-generation particles (whose angle is denoted as $\phi$) that are emitted independently according to the one-particle distribution Eq.~\eqref{oneParDis}, where the flow harmonics are determined.
Subsequently, one introduces an artificial decay process that occurs arbitrarily with a given probability $P_\mathrm{dcy}$ for a first-generation particle.
The chosen particle then decays into daughter particles, referred to as second-generation particles, according to a $1\to 2$ decay law.
Specifically, a mother particle is split into two daughter particles whose angular separation (a random variable denoted as $X$, with possible value $\varphi$) is described by a probability density function $g(\varphi)\equiv g(\varphi;\phi)$, such that $P(\varphi \leq X < \varphi + d\varphi) \approx g(\varphi)\, d\varphi$.
We assume that the random variable
\bqn
X \sim N\left(\phi, \sigma^2\right)  \label{GDisDecay}
\eqn
is normally distributed with mean $\phi$ and standard deviation $\sigma$.
The flow analysis is performed using the azimuthal angles $\varphi$ of the second-generation particles and the remaining undecayed first-generation particles.
This scheme is illustrated in the left panel of Fig.~\ref{fig_toymodel_scheme}.

The second toy model mimics the effect of momentum conservation.
While the decay process explored by toy model I produces localized particle pair correlations, the momentum conservation acts as a global constraint affecting all particles in the entire event.
While in practice the impact of momentum conservation on flow is usually negligible except at low multiplicity~\cite{hydro-corr-non-flow-01, hydro-corr-non-flow-07, hydro-corr-non-flow-14, hydro-corr-non-flow-20}, we intentionally amplify this effect in constructing our toy model.
To simplify our scenario, we consider that all the emitted particles have the same modulus $p_\mathrm{T}=1$.
To enhance the associated non-flow effect, we enforce momentum conservation for individual, smaller subsets of particles.
Specifically, an event is generated according to the following scheme.
One first generates a group of $M_\mathrm{sub}^\mathrm{min}$ (e.g., three) particles independently according to the one-particle distribution function Eq.~\eqref{oneParDis}.
One then evaluates the modulus of the total momentum $p_\mathrm{T}^\mathrm{total}$ of these particles. 
If the modulus falls within the interval $p_\mathrm{T}^\mathrm{total}\in \left[1, 2\right]$, one then generates two additional particles so that the total momentum vanishes.
Instead, if the total modulus falls in the interval $p_\mathrm{T}^\mathrm{total}\in \left[0, 1\right]$ or $\left[2, 3\right]$, one independently generates one more particle and falls back to the last step to check the total modulus until it falls into the desired interval $\left[1, 2\right]$.
Otherwise, while repeating the above process, if the total number of particles in the group keeps increasing and the modulus of the total momentum does not fall into the desired interval, one eventually casts away the group as the total number of particles exceeds a given value (e.g., seven), denoted as $M_\mathrm{sub}^\mathrm{max}$.
An event is furnished by a collection of such subsets of smaller multiplicities $M_\mathrm{sub}\in [M_\mathrm{sub}^\mathrm{min}, M_\mathrm{sub}^\mathrm{max}]$ using the above algorithm.
By breaking an event into smaller subsets that individually conserve the transverse momentum, we effectively enhance the resulting non-flow effect.
This scheme is illustrated in the right panel of Fig.~\ref{fig_toymodel_scheme}.

\section{Numerical results}\label{section4}

Based on the discussion in the previous section, we implemented two toy models to mimic non-flow effects corresponding to the particle decay and momentum conservation in realistic events. 
To assess the performance of the MLE method in suppressing non-flow contributions, we present the numerical results from both models below.
For comparison, we also include the results obtained by two approaches: particle correlation and the event-plane method.

In our MC-toy models, the first-generation particles are considered to be independently emitted, where the azimuthal distribution follows the one-particle probability density function given in Eq.~\eqref{oneParDis}.
For the numerical calculations, we retain harmonic contributions up to the fourth order while neglecting directed flow ($v_1$):
\begin{eqnarray}
f_1(\phi)=\frac{1}{2\pi }\left[1+2v_2\cos{2(\phi-\Psi_2)}+2v_3\cos{3(\phi-\Psi_3)}+2v_4\cos{4(\phi-\Psi_4)}\right] ,
\label{eqfphiv234}
\end{eqnarray}
where the event planes $\Psi_n$ are randomized for each event.

\subsection{MLE approach for toy model I}\label{sectionA}

As described in the previous section, an event is generated by considering an artificial decay process occurring to the first-generation particles.
Specifically, a fixed proportion of first-generation particles is randomly selected to decay.
A randomly chosen decaying particle is split into two daughter particles, emitted symmetrically around the original azimuthal angle $\phi$ with respect to the azimuthal direction of the mother particle, drawn from the Gaussian distribution Eq.~\eqref{GDisDecay}.
As a result, a generated event contains both decayed second-generation particles and undecayed first-generation particles.

To examine the performance of different flow estimation methods, we first show the results for a few randomly selected representative events.
For these events, the elliptic flow harmonics $v_2$ are sampled from a uniform distribution 
\bqn
X \sim U\left(0.2, 0.5\right) . \label{v2DisUni}
\eqn
The first-generation particles were generated independently according to Eq.~\eqref{eqfphiv234}, and the multiplicity is set to $M_0=3000$.
One considers that the probability for a first-generation particle to undergo the decay process is one third $P_\mathrm{dcy}=1/3$, which gives rise to roughly a multiplicity $M=4000$ per event.
The daughter particles follow a Gaussian distribution with a standard deviation of $\sigma=15^\circ$.

In Tab.~\ref{tabToy1sgl}, we compare the estimated $v_2$ using three different approaches, while comparing them with their input values.
In all five events, the results of the MLE method are closer to the input values, while all three approaches consistently underestimate the elliptic flow.
This difference becomes more significant for smaller $v_2$, which is observed as the number of events increases further.
The underestimation of flow harmonics is understood as the particle decay process effectively disrupts the statistical distribution generated by the one-particle distribution function Eq.~\eqref{oneParDis}.

\begin{table}
\setlength{\tabcolsep}{12pt}
\caption{The estimation of the elliptic flow $v_2$ from a few randomly generated events for toy model I.
The events are generated with a multiplicity of first-generation particles $M_0=3000$, $P_\mathrm{dcy}=1/3$, and $\sigma=15^\circ$.
The input value of the elliptic flow is drawn from the uniform distribution Eq.~\eqref{v2DisUni}.} \label{tabToy1sgl}
\begin{tabular}{c ccccc }
        \hline\hline
            method         & event 1 & event 2 & event 3 & event 4 & event 5 \\
          \hline
            input value            & 0.479 & 0.450 & 0.434 & 0.280 & 0.215\\
            MLE                    & 0.434 & 0.433 & 0.408 & 0.248 & 0.201\\
            event-plane            & 0.424 & 0.422 & 0.396 & 0.237 & 0.193\\
            particle correlation   & 0.423 & 0.421 & 0.395 & 0.236 & 0.191\\
         \hline\hline
\end{tabular}
\end{table}

In what follows, we evaluate the performance of different methods using more statistics, where the estimated flow harmonics $v_n$ are averaged over a large number of events.
In the calculations, we assume the values $v_2=0.2$, $v_3=0.15$, and $v_4=0.25$ in Eq.~\eqref{eqfphiv234}.

We first examine the impact on the estimation of $v_2$ due to the decay probability $P_\mathrm{dcy}$ and the width of the Gaussian distribution $\sigma$.
The numerical results are obtained using $N=1000$ events with multiplicity of the first-generation particles $M_0=3000$.
Tab.~\ref{tabToyratio} summarizes the estimations using different methods.
For the MLE and event-plane methods, one directly estimates the elliptic flow $v_2$ using the corresponding estimator $\widehat{v_2}$, and then takes the event average $\mu[\widehat{v_2}]$.
For the two-particle correlation method, however, the elliptic flow is estimated by taking the square root of the event average of the r.h.s. of Eq.~\eqref{eqEst2}, $\sqrt{\mu\left[\widehat{v_2^2}\right]}$ instead of $\mu\left[\sqrt{\widehat{v_2^2}}\right]$, in accordance with the practice~\cite{hydro-corr-ph-44}.
The top three rows compare results for increasing decay probability $P_\mathrm{dcy}$ at fixed width $\sigma=15^\circ$, while the bottom three rows show the effect of increasing $\sigma$ at fixed $P_\mathrm{dcy}=1/3$.
From Tab.~\ref{tabToyratio}, one observes that MLE performs slightly but consistently better across the entire table when compared to the two other methods.
It is understood that increasing the decay fraction or widening the decay angle distribution enhances the non-flow effect.
As a result, the estimated elliptic flow coefficient deviates further away from the input values.

\begin{table}
\setlength{\tabcolsep}{12pt}
\caption{The average estimation of the elliptic flow $v_2$ for toy model I using different methods.
The results are obtained using $N=1000$ events with a multiplicity of the first-generation particles $M_0=3000$, for different values of decay probability $P_\mathrm{dcy}$ and width $\sigma$ of the Gaussian distribution Eq.~\eqref{GDisDecay}.
The top three rows use a fixed width $\sigma=15^\circ$ with varying $P_\mathrm{dcy}$, while the bottom three rows use a fixed $P_\mathrm{dcy}=1/3$ with varying $\sigma$.}\label{tabToyratio}
\vspace{0.5em}
\begin{tabular}{c cccc}
\hline\hline
method    & particle correlation & event-plane & MLE & input value \\
\hline
\rule[-2ex]{0pt}{6ex}$P_\mathrm{dcy}$ & $\sqrt{\mu\left[\widehat{v_2^2}\right]}$ & $\mu\left[\widehat{v_2}\right]$ & $\mu\left[\widehat{v_2}\right]$ & $v_2$ \\ 
\hline
1/5  & 0.171 & 0.171 & 0.172 & 0.2 \\
1/3  & 0.157 & 0.157 & 0.158 & 0.2 \\
1/2  & 0.142 & 0.143 & 0.144 & 0.2 \\
\hline\hline
$\sigma$  &  &  &  &  \\
\hline
$\pi/24$  & 0.196 & 0.196 & 0.199 & 0.2 \\
$\pi/12$  & 0.189 & 0.190 & 0.191 & 0.2 \\
$\pi/6$   & 0.170 & 0.171 & 0.172 & 0.2 \\
\hline\hline
\end{tabular}
\end{table}

Now we turn to the results for different harmonics.
In our calculations, the width of the decay angle is set to $\sigma = 15^\circ$ and the decay probability is taken to be $P_\mathrm{dcy}=1/3$.
Besides, we consider different multiplicities for the first-generation particles $M_0 = 300$, $500$, $1000$, $3000$, and $6000$.
The numerical results are shown in Tab.~\ref{tabToyAvg}.
For the elliptic flow $v_2$, all methods slightly underestimate the input value, but the MLE gives the closest estimation.
A similar trend is observed for the quadrangular flow $v_4$.
In contrast, for the triangular flow $v_3$, the MLE estimation is slightly inferior compared to the other two methods.
This exception might be due to the mechanism of toy model I.
While disrupting the original distribution that suppresses the triangular flow, it also boosts $v_3$ by removing one particle (which can be effectively viewed as generating a particle in the opposite direction), meanwhile generating two particles with a splitting azimuthal angle.
Such a cancellation, somehow, undermines the precision of the estimation carried out using the likelihood function.
Overall, the performance of MLE is reasonable, consistent with existing approaches.

\begin{table}
\setlength{\tabcolsep}{12pt}
\caption{The averaged estimation of the flow harmonics $v_n$ for toy model I using different methods. 
The results are obtained using $N=1000$ events with different multiplicities $M_0$.
The averages $\mu [\cdots]$ of the estimators are evaluated on an event-by-event basis.}\label{tabToyAvg}
\vspace{0.5em}
\begin{tabular}{c cccc}
\hline\hline
method    & particle correlation & event-plane & MLE & input value \\
\hline
\rule[-2ex]{0pt}{6ex}$M_0$ & $\sqrt{\mu\left[\widehat{v_2^2}\right]}$ & $\mu\left[\widehat{v_2}\right]$ & $\mu\left[\widehat{v_2}\right]$ & $v_2$ \\ 
\hline
300   & 0.188 & 0.189 & 0.190 & 0.2 \\
500   & 0.187 & 0.187 & 0.188 & 0.2 \\
1000  & 0.188 & 0.188 & 0.189 & 0.2 \\
3000  & 0.186 & 0.186 & 0.187 & 0.2 \\
6000  & 0.186 & 0.186 & 0.187 & 0.2 \\
\hline\hline
\rule[-2ex]{0pt}{6ex} & $\sqrt{\mu\left[\widehat{v_3^2}\right]}$ & $\mu\left[\widehat{v_3}\right]$ & $\mu\left[\widehat{v_3}\right]$ & $v_3$ \\ 
\hline
300   & 0.130 & 0.134 & 0.133 & 0.15 \\
500   & 0.129 & 0.132 & 0.130 & 0.15 \\
1000  & 0.129 & 0.130 & 0.128 & 0.15 \\
3000  & 0.130 & 0.130 & 0.128 & 0.15 \\
6000  & 0.130 & 0.130 & 0.127 & 0.15 \\
\hline\hline
\rule[-2ex]{0pt}{6ex} & $\sqrt{\mu\left[\widehat{v_4^2}\right]}$ & $\mu\left[\widehat{v_4}\right]$ & $\mu\left[\widehat{v_4}\right]$ & $v_4$ \\ 
\hline
300   & 0.197 & 0.199 & 0.199 & 0.25 \\
500   & 0.196 & 0.197 & 0.197 & 0.25 \\
1000  & 0.195 & 0.196 & 0.196 & 0.25 \\
3000  & 0.196 & 0.196 & 0.196 & 0.25 \\
6000  & 0.196 & 0.196 & 0.196 & 0.25 \\
\hline\hline
\end{tabular}
\end{table}

\subsection{Revised MLE for toy model I}\label{sectionA2}

Up to this point, we have shown that MLE provides a reasonable estimation for flow harmonics.
Compared to existing approaches, it exhibits a certain degree of suppression for the non-flow caused by the decay process; however, its performance is not significantly improved.
This is partly because the likelihood employed by MLE is governed by the one-particle distribution function Eq.~\eqref{oneParDis}, which does not take into account any particular information regarding the decay process.
Moreover, the particle decay introduced in toy model I involves a finite fraction $P_\mathrm{dcy}$ of total multiplicity. 
In other words, the process is more of a collective phenomenon, and will not be easily suppressed when one resorts to high-order correlators, as would be expected from a non-flow origin that scales inversely with the multiplicity as suggested by Borghini {\it et al.}~\cite{hydro-corr-ph-03, hydro-corr-ph-04}.

Based on the above observation, it is therefore natural to ask whether it would be possible to further improve the estimation scheme by explicitly taking into account available information.
This can be elaborated as follows.
One assumes that an unknown fraction of the first-generation particles undergoes the decay process, satisfying a distribution $g(\varphi;\phi)$ by Eq.~\eqref{GDisDecay}.
Specifically, the parameter space of the MLE is expanded to
\bqn
\theta=\left(v_1, v_2, \cdots, \Psi_1, \Psi_2, \cdots, P_\mathrm{dcy},\sigma\right) .
\eqn
Besides the flow harmonics, the unknown parameters related to the specific physical scenario have also been taken into account.
This flexibility is an interesting feature of MLE and does not straightforwardly generalize to other flow estimation schemes.

For an event of multiplicity $M$ the likelihood Eq.~\eqref{eqlikelihood} now becomes
\begin{eqnarray}
\mathcal{L}(\theta; \varphi_{1}, \cdots, \varphi_M) 
= &&\bigcup_{M_\mathrm{pair}=0}^{\left[M/2\right]} \left(1-P_\mathrm{dcy}\right)^{M-2M_\mathrm{pair}}\bigcup_{\mathrm{pairings}}
\mathcal{L}_\mathrm{nodcy}(\theta;\varphi_{1},\cdots, \varphi_{M-2M_\mathrm{pair}}) \nb\\
&\times&\prod_{j=1}^{M_\mathrm{pair}} P_\mathrm{dcy}
\cdot f_1\left(\frac{\varphi_{j_1}+\varphi_{j_2}}{2}; \theta\right)
\cdot g\left(\varphi_{j_1}-\varphi_{j_2}; \frac{\varphi_{j_1}+\varphi_{j_2}}{2}\right) ,
\label{eqlikelihoodPNew}
\end{eqnarray}
where
\begin{eqnarray}
\mathcal{L}_\mathrm{nodcy}(\theta; \varphi_{1}, \cdots, \varphi_M) =\prod_{k=1}^{M}f_1(\varphi_k; \theta) \nb \label{eqlikelihoodNN}
\end{eqnarray}
for undecayed $M-2M_\mathrm{pair}$ first-generation particles is essentially given by Eq.~\eqref{eqlikelihood}.
We note that the subscript ``pairings'' implies all possible ways of picking out $2M_\mathrm{pair}$ particles and combining them into different $M_\mathrm{pair}$ pairs, where the combination number $\binom{M}{M_0}(2M_\mathrm{pair}-1)!!$ is implied and $j_1$ and $j_2$ are the two second-generation particles constituting the $j$th pair.

In practice, exhaustively evaluating Eq.~\eqref{eqlikelihoodPNew}, which is already cumbersome in form, is highly time-consuming and hardly feasible on a personal computer.
A compromised approach is to approximate the likelihood by introducing some educated simplification.
Firstly, the number of emitted pairs is expected to stay close to the value 
\bqn
M_\mathrm{pair}\approx  \frac{P_\mathrm{dcy}}{1+P_\mathrm{dcy}}M \approx P_\mathrm{dcy} M_0 \ .
\eqn
One does not need to enumerate values of $M_\mathrm{pair}$ that deviate significantly.
Secondly, we will selectively pick up a small fraction of pair combinations that are deemed more probable.
This can be done by either summing up the angle difference and only considering a fraction of combinations that have the smallest sum, or using an attentive value of $\sigma=\sigma_E$ (in our calculations, we randomly choose $\sigma_E$ from the interval $[0.15, 0.45]$) for $g(\varphi)$ to reject most less probable pairs from the simulated data.
The obtained likelihood is then minimized to obtain the flow harmonics, as described above.

\begin{table}[ht]
\setlength{\tabcolsep}{10pt}
\caption{The average estimation of the elliptic flow $v_2$ and decay parameters $\sigma$ and $P_\mathrm{dcy}$ using the revised MLE approach for events with different input values.
The calculations are carried out for $N=1000$ events with $M_0=2000$ and for the specific scenario $P_\mathrm{dcy}=1$.}\label{tabV2SigmaFit}
\begin{tabular}{cccccc}
\hline\hline
$v_2$  & $v_2$ (input value) & $\sigma$  & $\sigma$ (true value)& $P_\mathrm{dcy}$  & $P_\mathrm{dcy}$ (true value) \\
\hline
0.146 & 0.15 & 0.204 & $\pi/12$ & 0.925 & 1 \\
0.192 & 0.20 & 0.209 & $\pi/12$ & 0.964 & 1\\
0.283 & 0.30 & 0.205 & $\pi/12$ & 0.988 & 1\\
0.375 & 0.40 & 0.206 & $\pi/12$ & 0.935 & 1 \\
\hline\hline
\end{tabular}
\end{table}

\begin{table}[ht]
\setlength{\tabcolsep}{10pt}
\caption{The average estimation of the elliptic flow $v_2$ using different methods for events with different multiplicities of the first-generation particles $M_0$.
The calculations are carried out for $N=1000$ events and for the specific scenario $P_\mathrm{dcy}=1$.}\label{tabCompareMethods}
\begin{tabular}{c c c c c c}
\hline\hline
$M_0$ & particle correlation & event-plane & MLE & rev-MLE  & input value\\
\hline
500  & 0.181 & 0.180 & 0.185 & 0.195 & 0.2\\
1000  & 0.175 & 0.174 & 0.180 & 0.191 & 0.2\\
2000  & 0.175 & 0.176 & 0.179 & 0.191 & 0.2\\
3000  & 0.174 & 0.173 & 0.177 & 0.189 & 0.2 \\
\hline\hline
\end{tabular}
\end{table}

The numerical results are shown in Tabs.~\ref{tabV2SigmaFit} and~\ref{tabCompareMethods}.
By using the revised MLE approach, we show the event average of estimated elliptic flow $v_2$, decay angular width $\sigma$, and probability $P_\mathrm{dcy}$ in Tab.~\ref{tabV2SigmaFit} for simulated data generated using different input values of $v_2$.
The calculations are carried out for events with the multiplicity of the first-generation particles $M_0=2000$, and the event average is carried out for $N=1000$ events.
Compared to previous results, it is observed that the estimated $v_2$ converges much better to its input values, even though the estimations of $\sigma$ and $P_\mathrm{dcy}$ are not perfect owing to the adopted approximations.

Tab.~\ref{tabCompareMethods} summarizes the estimated elliptic flow $v_2$ using different methods.
The calculations are carried out for events with different multiplicities $M_0$, and the event average is carried out for $N=1000$ events.
While the results obtained by particle correlation, event-plane, and MLE methods underestimate $v_2$ across all multiplicities, the revised MLE brings the estimation much closer to the input values.
For instance, at $M_0=500$, the estimated elliptic flow obtained by the revised MLE improves by approximately $7.7\%$ over the particle correlation method and by about $5.4\%$ over the original MLE.
The improved performance indicates that when it is feasible for the MLE approach to expand its parameter space appropriately, the improvement might be substantial.

\subsection{A comparison between toy model I and existing studies}\label{sectionA3}

The impact of particle emission on collective flow has been explored through numerical simulations in the literature~\cite{hydro-corr-ph-03, hydro-corr-ph-04, RHIC-star-v2-07}. 
In~\cite{hydro-corr-ph-03, hydro-corr-ph-04}, additional particle pairs with identical azimuthal angles are emitted on top of the background flow. 
In a further analysis~\cite{RHIC-star-v2-07} conducted by the STAR collaboration, the emitted particle pairs are separated by a fixed angle. 
The resulting elliptic flow is evaluated as a function of the opening angle using various approaches. 
In all these studies, the estimated flow exceeds the true value when the opening angle of the emitted particle pairs is small or vanishing. 
This behavior seems to be precisely the opposite of what was found in the previous subsections, where the input value was underestimated. 
We therefore owe an explanation for the subtle differences between the models that lead to these distinct results. 

Before presenting the numerical results, let us first elaborate on the interplay between several competing factors. 
\begin{itemize}
\item Opening angle of the emitted particle pairs: small or back-to-back angles versus near 90 degrees; 
\item Order of the correlator: two-particle versus four-particle correlators; 
\item Correlation with the event plane: correlated versus uncorrelated emissions. 
\end{itemize} 
It is noted that the true values of the flow harmonics, defined as~\cite{RHIC-star-v2-07}
\bqn
v_n^\mathrm{true}\equiv \langle\cos n(\phi-\Psi_n)\rangle ,\label{vnTrue}
\eqn
can be analytically derived for a fixed opening angle $\phi_\mathrm{open}$.  
Consider $M_\mathrm{pair}$ pairs emitted in an event of total multiplicity $M$.  
If the pair emission is uncorrelated with the symmetry plane, we have
\bqn
v_n^\mathrm{true} = \frac{(M-2M_\mathrm{pair})v_n}{M} .\label{vnUncorrTrue}
\eqn
Instead, if the pair emission is correlated with the symmetry plane, we obtain
\bqn
v_n^\mathrm{true} = \left[\frac{M-M_\mathrm{pair}}{M}+ \cos (n\phi_\mathrm{open})\frac{M_\mathrm{pair}}{M}\right]v_n ,\label{vnCorrTrue}
\eqn
which yields $v_2^\mathrm{true} = v_2$ when the particle pairs are either back-to-back or perfectly aligned.

For both the two-particle and four-particle correlators, the fraction of combinations effectively impacted by pair emission increases with the number of pairs.  
However, the relative contribution is much smaller for the four-particle correlator.  
For perfectly aligned and back-to-back particle pairs whose emission is correlated with the symmetry plane, both correlators are expected to overestimate the input value of elliptic flow, which, in this specific case, coincides with the true value according to Eq.~\eqref{vnCorrTrue}.  
This effect is minor for the four-particle correlator, especially as the number of emitted particle pairs increases, as illustrated in Tab.~1 of~\cite{hydro-corr-ph-04} and Fig.~9 of~\cite{RHIC-star-v2-07}.  
For a fixed number of particle pairs, when the emission is correlated with the symmetry plane and the opening angle is varied, the extracted elliptic flow exceeds the input value at small and back-to-back angles, but falls below it near $90^\circ$ for both correlators.  
The latter suppression is more pronounced and is naturally understood, since particles emitted out of plane tend to reduce elliptic flow.  
Moreover, the elliptic flow values obtained from the two-particle correlator do not differ significantly from those associated with the four-particle correlator.  
If, instead, the pair emission is uncorrelated with the symmetry plane, the estimated flow fluctuates around the true value given by Eq.~\eqref{vnUncorrTrue} and is therefore systematically smaller than the input value.  
This behavior is intuitive: the non-flow contribution is isotropic by construction, reducing the apparent flow signal, and the degree of suppression depends on the opening angle. 
As shown in Fig.~11 of~\cite{RHIC-star-v2-07} and discussed below, the four-particle correlator $v_2\{4\}$ provides an almost unbiased estimate of the true value, whereas $v_2\{2\}$ exhibits more pronounced oscillations and underestimates the flow for opening angles around $90^\circ$.  

\begin{figure}[ht]
    \centering
    \begin{minipage}{0.4\textwidth}
        \centering
        \includegraphics[width=1.2\textwidth, height=0.3\textheight]{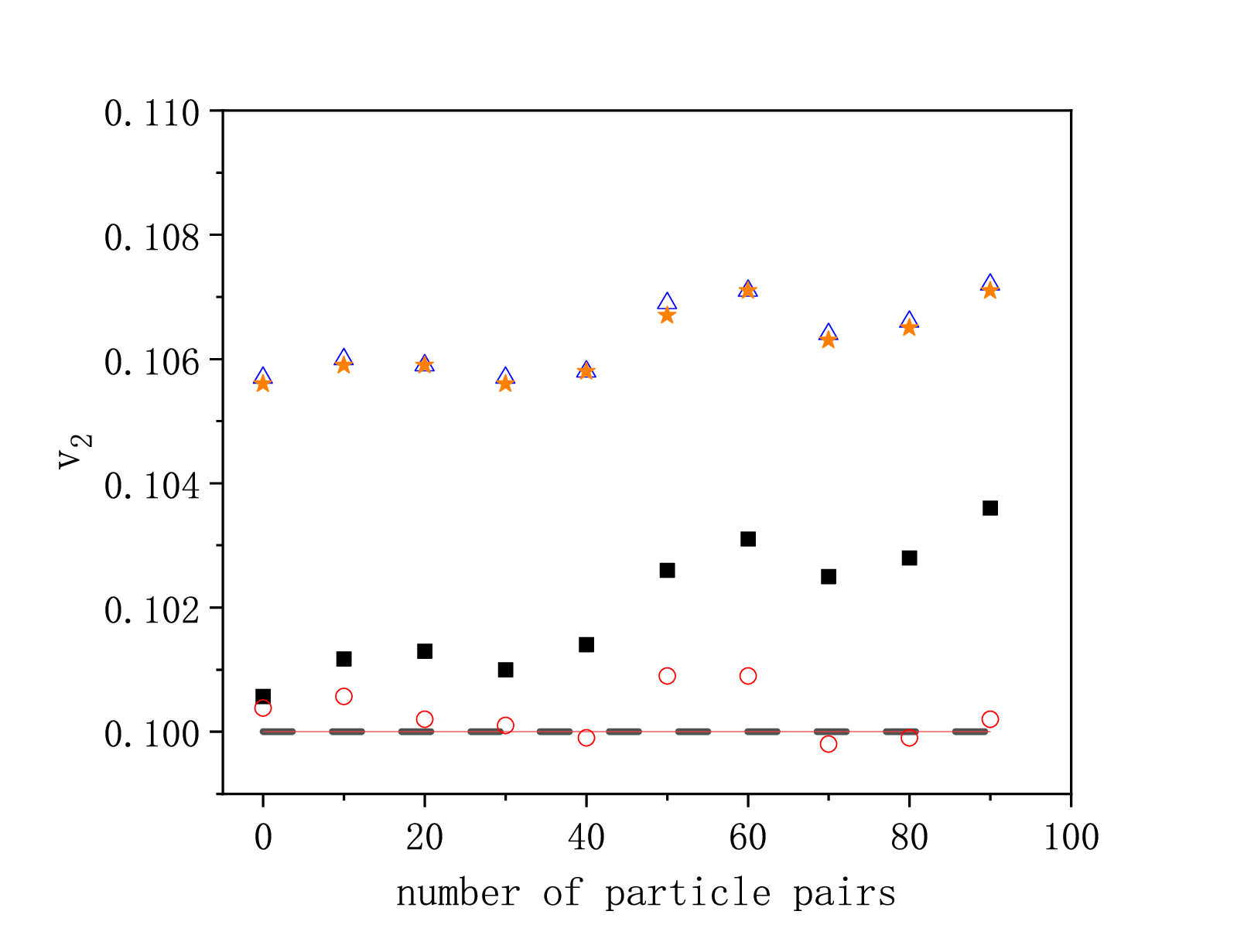}
    \end{minipage}
    \begin{minipage}{0.4\textwidth}
        \centering
        \includegraphics[width=1.2\textwidth, height=0.3\textheight]{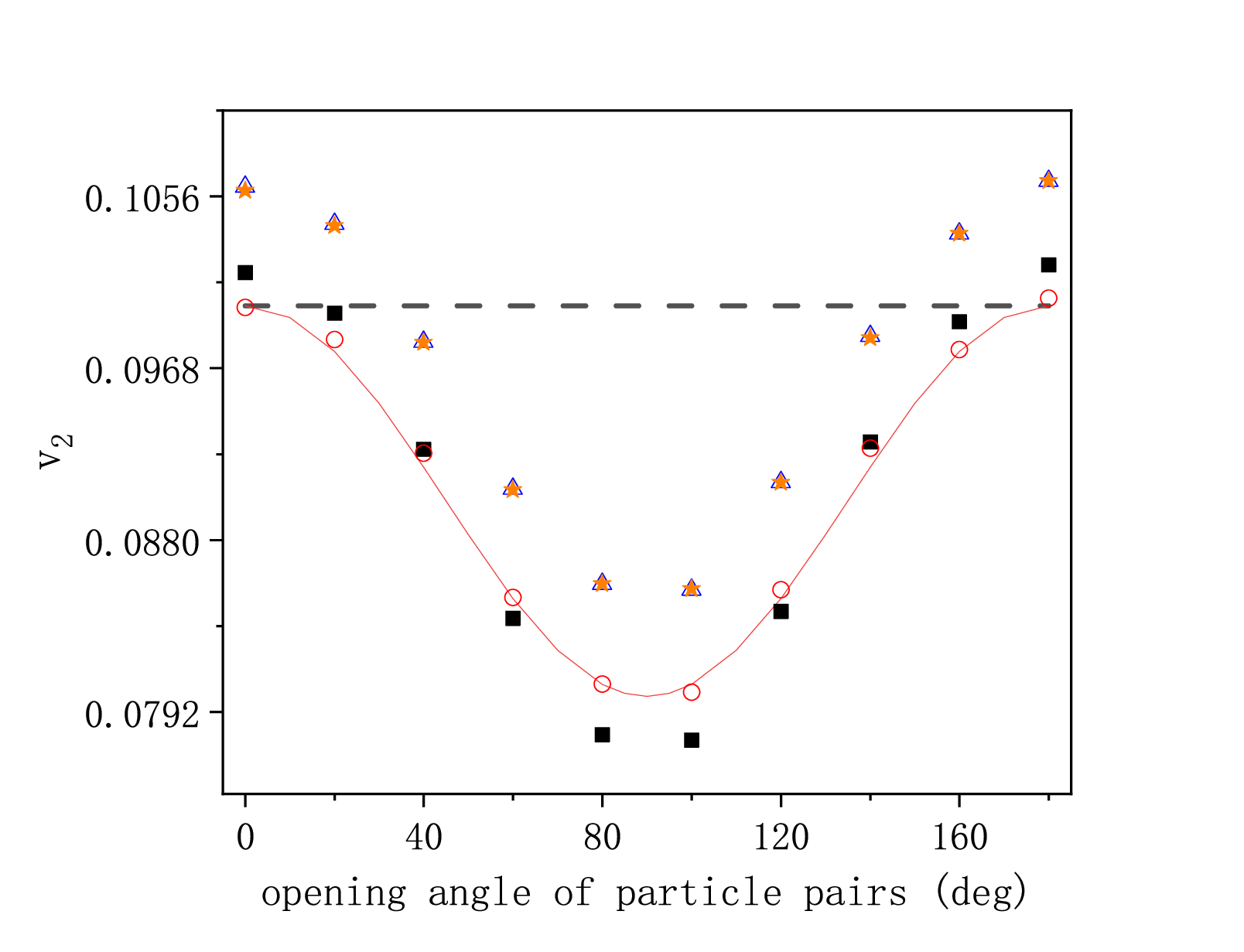}
    \end{minipage}
    \begin{minipage}{0.4\textwidth}
        \centering
        \includegraphics[width=1.2\textwidth, height=0.3\textheight]{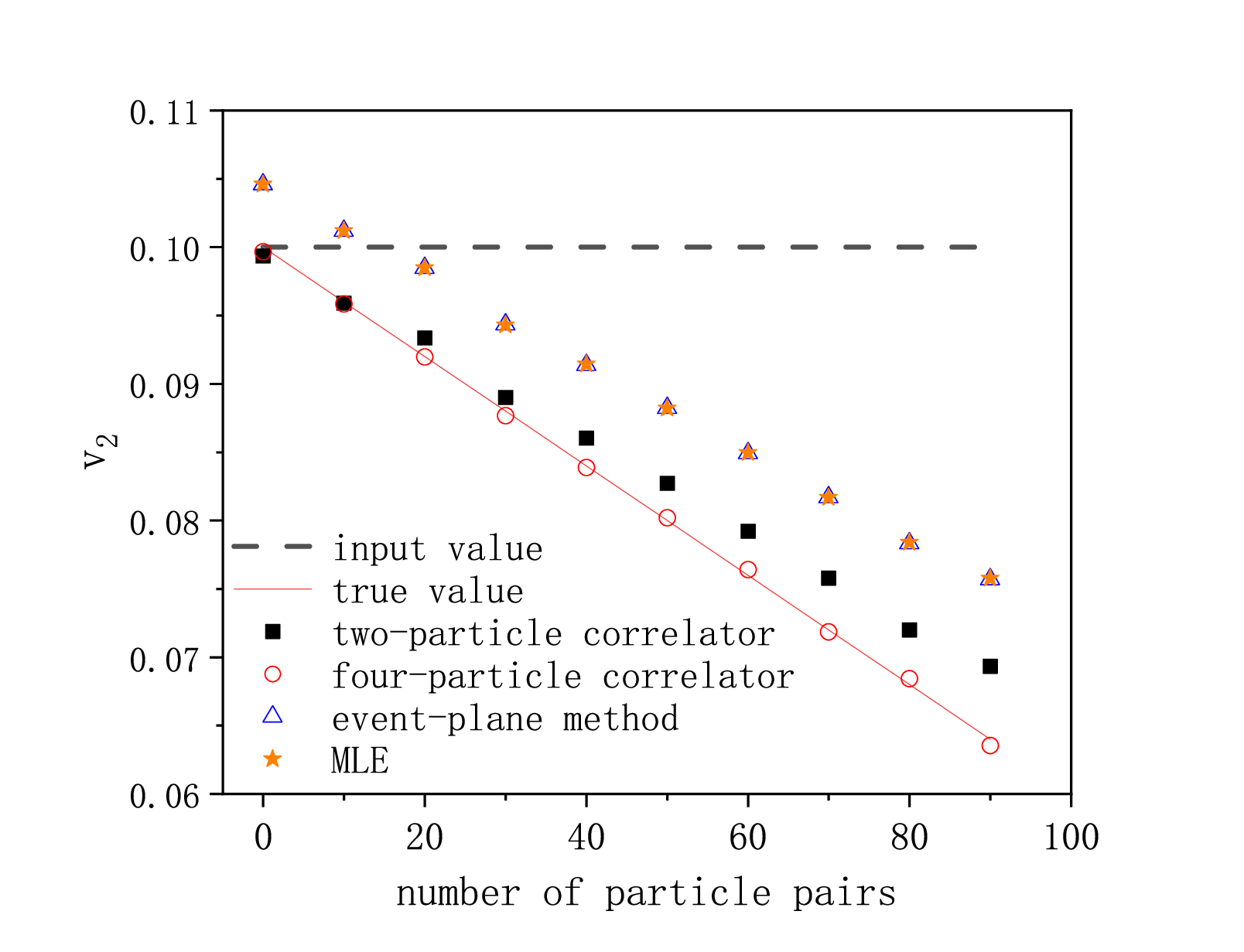}
    \end{minipage}
    \begin{minipage}{0.4\textwidth}
        \centering
        \includegraphics[width=1.2\textwidth, height=0.3\textheight]{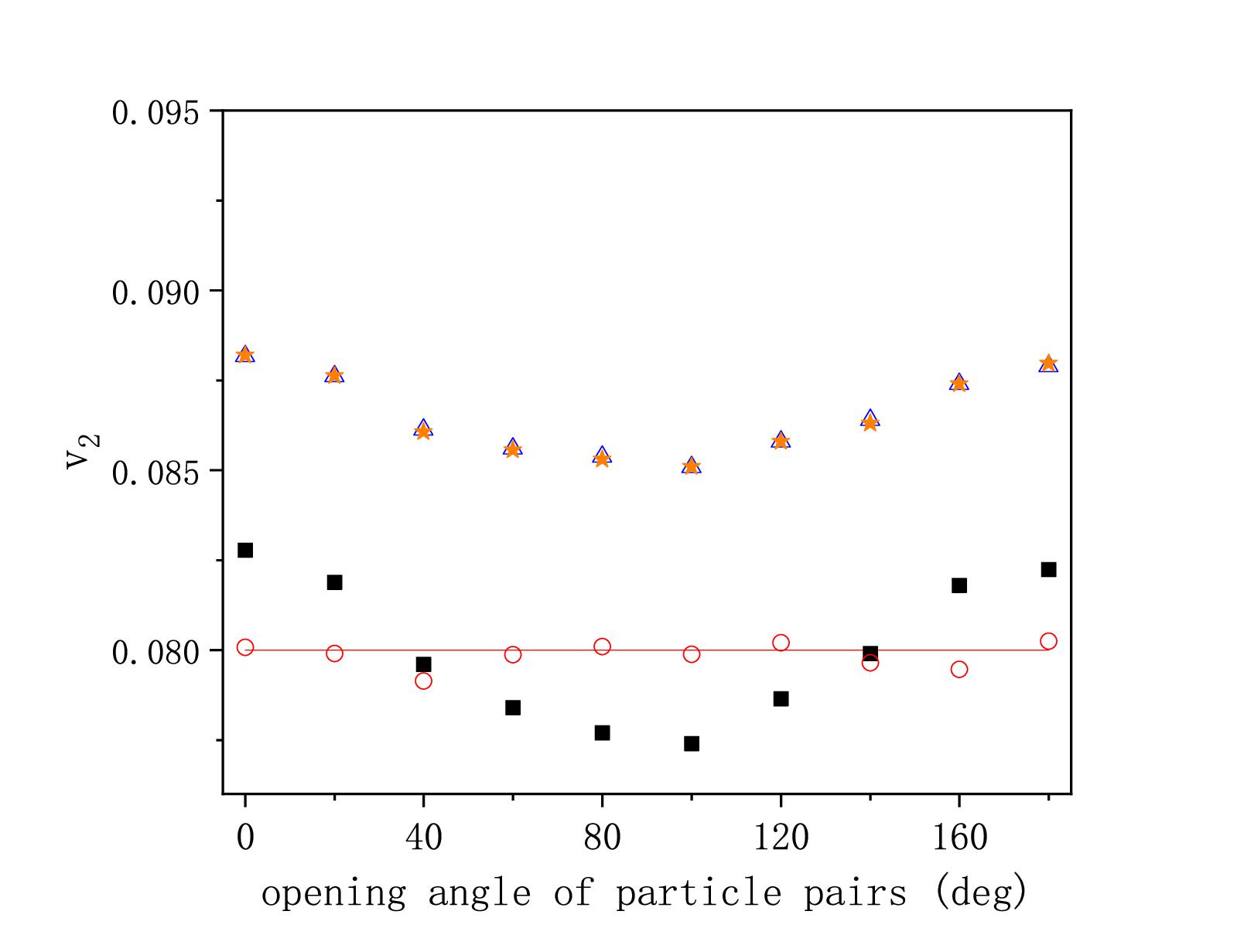}
    \end{minipage}
\renewcommand{\figurename}{Fig.}   
\caption{Flow harmonics evaluated using particle correlators, the event-plane method, and MLE.
The events are generated from the emission of particle pairs with a prescribed opening angle, superimposed on a background collective flow $v_2 = 0.1$.
The input flow harmonics, where applicable, are indicated by a dashed black horizontal line, while the true values governed by Eqs.~\eqref{vnUncorrTrue} and~\eqref{vnCorrTrue} are shown by a thin red line.
The analysis is performed for 10 000 events, each of which contains a total of 500 particles.
Left column: The particle pairs are back-to-back, and the elliptic flow is evaluated as a function of the number of particle pairs.
Right column: The number of particle pairs is fixed to 50, and the elliptic flow is evaluated as a function of the opening angle.
Top row: The particle pairs are correlated with the symmetry plane, with one particle of each pair emitted according to the background one-particle distribution.
Bottom row:  same as the top row, but the particle pairs are uncorrelated with the symmetry plane.}
\label{cmp_scheme1}
\end{figure}

Our numerical results are summarized in Fig.~\ref{cmp_scheme1}.  
First, we reproduce the calculations shown in Figs.~9 and~11 of the STAR paper~\cite{RHIC-star-v2-07}, displayed in the top-left and bottom-right panels of Fig.~\ref{cmp_scheme1}.  
We perform two sets of calculations.  

On the one hand, we consider the emission of back-to-back particle pairs and compute the flow harmonics as functions of the number of pairs.  
These back-to-back pairs are taken to be either uncorrelated or correlated with the symmetry plane, and the results are presented in the left column of Fig.~\ref{cmp_scheme1}.  
In the top-left panel, we show $v_2\{2\}$ and $v_2\{4\}$ as functions of the number of back-to-back pairs for events in which the pair emission is correlated with the symmetry plane, finding good agreement with the results reported in~\cite{RHIC-star-v2-07}.  
Both estimators overestimate the input value, and their deviations grow with the number of pairs, although the deviation associated with $c_2\{4\}$ remains much smaller than that of $c_2\{2\}$.  
We also include the results obtained with the MLE and event-plane methods, which likewise show an overestimation.  
In the bottom-left panel, we show the corresponding results when the pair emission is uncorrelated with the symmetry plane.  
In this case, $v_2\{2\}$ and $v_2\{4\}$ lie below the input value but remain close to the true value given by Eq.~\eqref{vnUncorrTrue}, for which the deviation from the input value increases with the number of particle pairs.  

On the other hand, for a fixed multiplicity, we evaluate the relevant observables as the opening angle is varied from $0$ to $\pi$, again considering both correlated and uncorrelated emission with respect to the event plane.  
These results are presented in the right column of Fig.~\ref{cmp_scheme1}.  
In the top-right panel, we display $v_2\{2\}$ and $v_2\{4\}$, which show good agreement with the true value given in Eq.~\eqref{vnCorrTrue}.  
As discussed above, $v_2\{2\}$ slightly exceeds the true value at small and back-to-back angles but falls below it near $90^\circ$, while the agreement for $v_2\{4\}$ is almost exact. 
The same qualitative behavior is observed for the MLE and event-plane methods, which yield values above the true one yet below the input value near $90^\circ$, consistent with the findings in the preceding subsection.
In the bottom-right panel, we show the results for events in which the pair emission is uncorrelated with the symmetry plane.  
The values of $v_2\{2\}$ and $v_2\{4\}$ can be directly compared with Fig.~11 of~\cite{RHIC-star-v2-07}; both estimators reproduce the true value reasonably well, with $v_2\{4\}$ exhibiting noticeably fewer oscillations.

\subsection{MLE approach for toy model II}\label{sectionB}

We now turn our attention to another important source of non-flow: global momentum conservation.
As mentioned, the model has been constructed in a way that the effect of momentum conservation becomes more pronounced.
This is accomplished by forcing momentum conservation for individual, smaller subsets of particles that constitute the event.

Our numerical results are shown in Tabs.~\ref{tabToy2sgl},~\ref{tabToy2Avg}, and~\ref{tabToyratio2}.
Tab.~\ref{tabToy2sgl} illustrates the elliptic flow $v_2$ extracted from five randomly selected events generated by toy model II.
We consider events of total multiplicity $M\approx 2400$ constituted by smaller subsets of multiplicity $M_\mathrm{sub}\in [5, 10]$.
Again, the input value of the elliptic flow is sampled from the uniform distribution Eq.~\eqref{v2DisUni}.
It is noted that the total multiplicity $M$ is not strictly fixed owing to the model's specific construction.
By comparing the results of different approaches with the input value, all three methods underestimate $v_2$, indicating an overall suppression effect on the collective flow resulting from momentum conservation.
Again, among the three methods, the estimate given by MLE is the closest to the input value.
As shown below, this result holds for various multiplicities, flow harmonics, and model configurations.

By employing more statistics, as shown in Tabs.~\ref{tabToy2Avg} and~\ref{tabToyratio2}, the calculations are carried out considering $N=1000$ events.
Tab.~\ref{tabToy2Avg} shows the estimations of different harmonics for events with different multiplicities using different approaches.
For the chosen model configuration $M_\mathrm{sub}\in [5, 10]$, the three different multiplicities correspond to roughly $100$, $200$, and $300$ subsets of particles.
As shown in Tab.~\ref{tabToy2Avg}, the MLE method consistently outperforms other methods for all three harmonics across different multiplicity settings.

We also attempt to explore the effectiveness of the methods concerning the impact of momentum conservation, which is achieved by tuning the model configuration through the size of the subset $M_\mathrm{sub}$.
It is understood that the impact of non-flow becomes more significant as the subset multiplicity $M_\mathrm{sub}$ decreases.
In Tab.~\ref{tabToyratio2}, we analyze the elliptic flow $v_2$ while varying the interval $[M_\mathrm{sub}^\mathrm{min}, M_\mathrm{sub}^\mathrm{max}]$.
As expected, it can be seen that as the subset multiplicity $M_\mathrm{sub}$ decreases, the estimation of the elliptic flow consistently deviates further from the input value.
This reflects that the non-flow correlations caused by momentum conservation gradually strengthen as the number of subsets increases.
When compared to other methods, the estimations made by the MLE approach stay closer to the input value, showing superior non-flow suppression ability.

\begin{table}
\setlength{\tabcolsep}{12pt}
\caption{The estimation of the elliptic flow $v_2$ from a few randomly generated events for toy model II.
The events are generated with a multiplicity $M\approx 2400$ and $M_\mathrm{sub}\in [5, 10]$.
The input value of the elliptic flow is drawn from the uniform distribution Eq.~\eqref{v2DisUni}.}\label{tabToy2sgl}
\begin{tabular}{c ccccc }
        \hline\hline
            method         & event 1 & event 2 & event 3 & event 4 & event 5 \\
          \hline
            input value             & 0.317 & 0.358 & 0.311 & 0.312 & 0.453\\
            MLE                    & 0.237 & 0.250 & 0.215 & 0.228 & 0.351\\
            event-plane            & 0.235 & 0.247 & 0.203 & 0.228 & 0.346\\
            particle correlation   & 0.235 & 0.246 & 0.202 & 0.227 & 0.346\\
         \hline\hline
\end{tabular}
\end{table}

\begin{table}
\setlength{\tabcolsep}{12pt}
\caption{The averaged estimation of flow harmonics $v_n$ for toy model II using different methods.
The results are obtained using $N=1000$ events with $M_\mathrm{sub}\in [5, 10]$ for different total multiplicities $M$.
The averages $\mu[\cdots]$ of the estimators are evaluated on an event-by-event basis.}\label{tabToy2Avg}
\vspace{0.5em}
\begin{tabular}{c cccc}
\hline\hline
method    & particle correlation & event-plane & MLE & input value \\
\hline
\rule[-2ex]{0pt}{6ex}$M$ & $\sqrt{\mu\left[\widehat{v_2^2}\right]}$ & $\mu\left[\widehat{v_2}\right]$ & $\mu\left[\widehat{v_2}\right]$ & $v_2$ \\ 
\hline
300   & 0.132 & 0.138 & 0.144 & 0.2 \\
500   & 0.131 & 0.135 & 0.141 & 0.2 \\
800   & 0.132 & 0.134 & 0.138 & 0.2 \\
1500  & 0.132 & 0.132 & 0.138 & 0.2 \\
2300  & 0.133 & 0.133 & 0.135 & 0.2 \\
\hline\hline
\rule[-2ex]{0pt}{6ex} & $\sqrt{\mu\left[\widehat{v_3^2}\right]}$ & $\mu\left[\widehat{v_3}\right]$ & $\mu\left[\widehat{v_3}\right]$ &  \\ 
\hline
300   & 0.100 & 0.108 & 0.113 & 0.15 \\
500   & 0.100 & 0.104 & 0.108 & 0.15 \\
800   & 0.100 & 0.103 & 0.108 & 0.15 \\
1500  & 0.100 & 0.102 & 0.106 & 0.15 \\
2300  & 0.100 & 0.101 & 0.109 & 0.15 \\
\hline\hline
\rule[-2ex]{0pt}{6ex} & $\sqrt{\mu\left[\widehat{v_4^2}\right]}$ & $\mu\left[\widehat{v_4}\right]$ & $\mu\left[\widehat{v_4}\right]$ &  \\ 
\hline
300   & 0.157 & 0.162 & 0.168 & 0.25 \\
500   & 0.157 & 0.160 & 0.166 & 0.25 \\
800   & 0.158 & 0.160 & 0.164 & 0.25 \\
1500  & 0.158 & 0.159 & 0.167 & 0.25 \\
2300  & 0.157 & 0.158 & 0.167 & 0.25 \\
\hline\hline
\end{tabular}
\end{table}

\begin{table}
\setlength{\tabcolsep}{12pt}
\caption{The averaged estimation of the elliptic flow $v_2$ for toy model II with different sizes of subsets using different methods.
The results are obtained using $N=1000$ events with given multiplicity $M\approx 2400$.}\label{tabToyratio2}
\begin{tabular}{c cccc}
\hline\hline
\rule[-2ex]{0pt}{5ex}$[M_\mathrm{sub}^\mathrm{min}, M_\mathrm{sub}^\mathrm{max}]$  & particle correlation & event-plane & MLE & input value \\
\hline
$[3, 7]$    & 0.115 & 0.116 & 0.124 & 0.2 \\
$[5, 10]$   & 0.132 & 0.132 & 0.140 & 0.2 \\
$[10, 15]$  & 0.149 & 0.149 & 0.156 & 0.2 \\
$[15, 20]$  & 0.154 & 0.153 & 0.164 & 0.2 \\
\hline\hline
\end{tabular}
\end{table}

\subsection{A comparison between toy model II and existing studies}\label{sectionB2}

As mentioned, the impact of global momentum conservation on flow and particle correlations has been investigated by many authors~\cite{hydro-corr-non-flow-01, hydro-corr-non-flow-02, hydro-corr-non-flow-03, hydro-corr-non-flow-07, hydro-corr-non-flow-08, hydro-corr-non-flow-13, hydro-corr-non-flow-14, hydro-corr-non-flow-20}. 
From an analytical perspective, these studies evaluate the correction to the $k$-particle correlation under the assumption of independent particle emission, using an identical one-particle distribution function in which transverse momentum conservation is imposed as a constraint. 
The calculation simplifies if one assumes an isotropic background, neglecting the harmonic coefficients of the flow. 
At the lowest order, the correction is non-vanishing only for the directed flow $v_1\{2\}= \sqrt{c_1\{2\}}$, where ${c_1\{2\}}\propto 1/M$ for a one-particle distribution with vanishing collective flow~\cite{hydro-corr-non-flow-01, hydro-corr-non-flow-02}. 
This result has been generalized~\cite{hydro-corr-non-flow-13, hydro-corr-non-flow-20} to yield the non-vanishing leading contribution $c_n\{2k\}\propto 1/(M-2k)^{nk}$. 
When the background flow is included, although generally small in magnitude, the corrections to the collective flows become more significant and depend sensitively on the type of correlator. 
The corrections to the correlators follow $\Delta c_2\{2\}\propto 1/M$ and $\Delta c_2\{4\}\propto 1/M$~\cite{hydro-corr-non-flow-14}, which appear to generalize to $\Delta c_n\{2k\}\propto 1/(M-2k)$~\cite{hydro-corr-non-flow-20}. 
In particular, it has been pointed out~\cite{hydro-corr-non-flow-14} that there is a competition between the contributions from collective flow and non-flow, so that $c_2\{4\}$ may change sign as the system's multiplicity varies. 
Nevertheless, the non-flow corrections remain positive for both $c_2\{2\}$ and $c_2\{4\}$. 

\begin{figure}[ht]
    \centering
    \begin{minipage}{0.4\textwidth}
        \centering
        \includegraphics[width=1.2\textwidth, height=0.3\textheight]{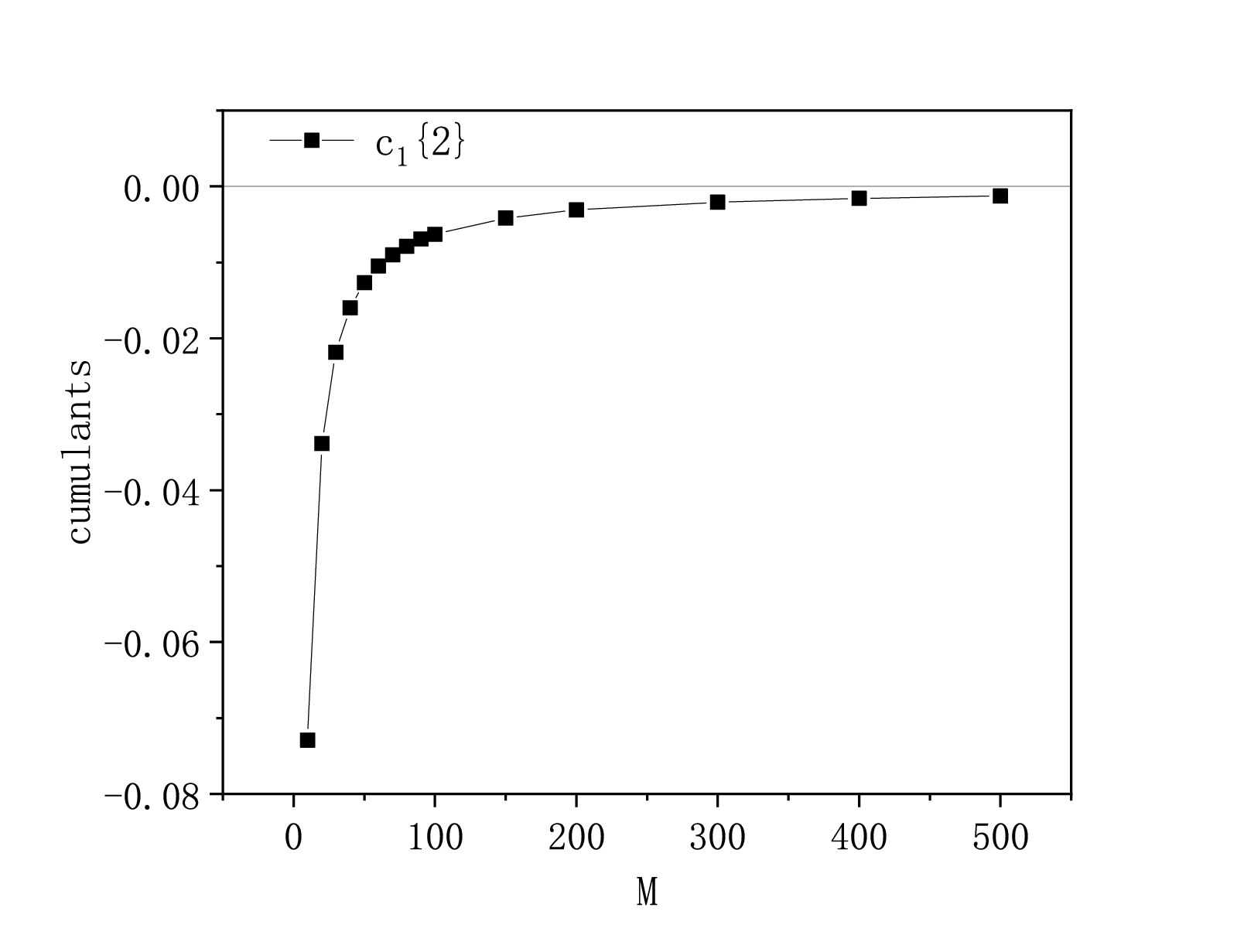}
    \end{minipage}
    \begin{minipage}{0.5\textwidth}
        \centering
        \includegraphics[width=1.0\textwidth, height=0.3\textheight]{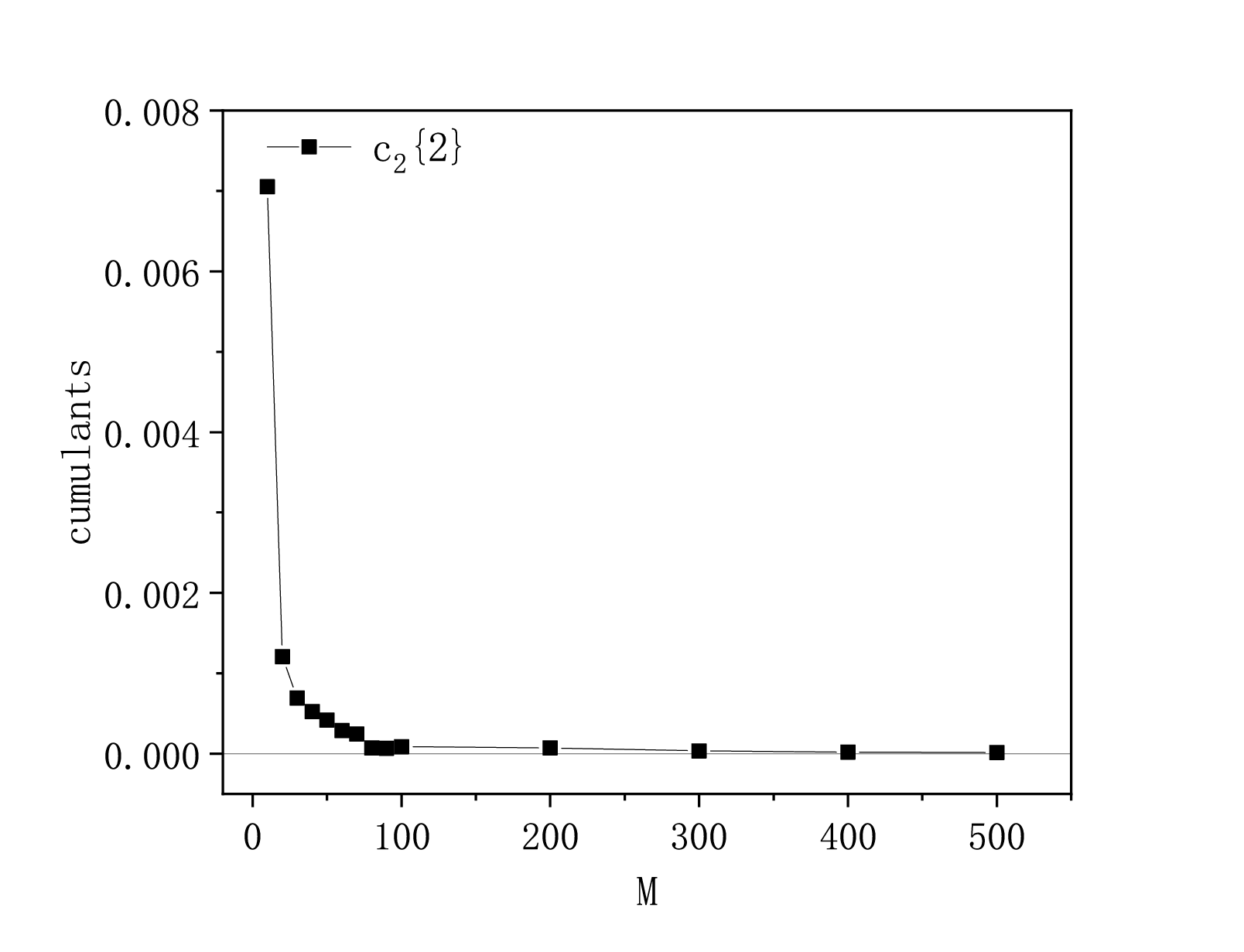}
    \end{minipage}
    \begin{minipage}{0.4\textwidth}
        \centering
        \includegraphics[width=1.2\textwidth, height=0.3\textheight]{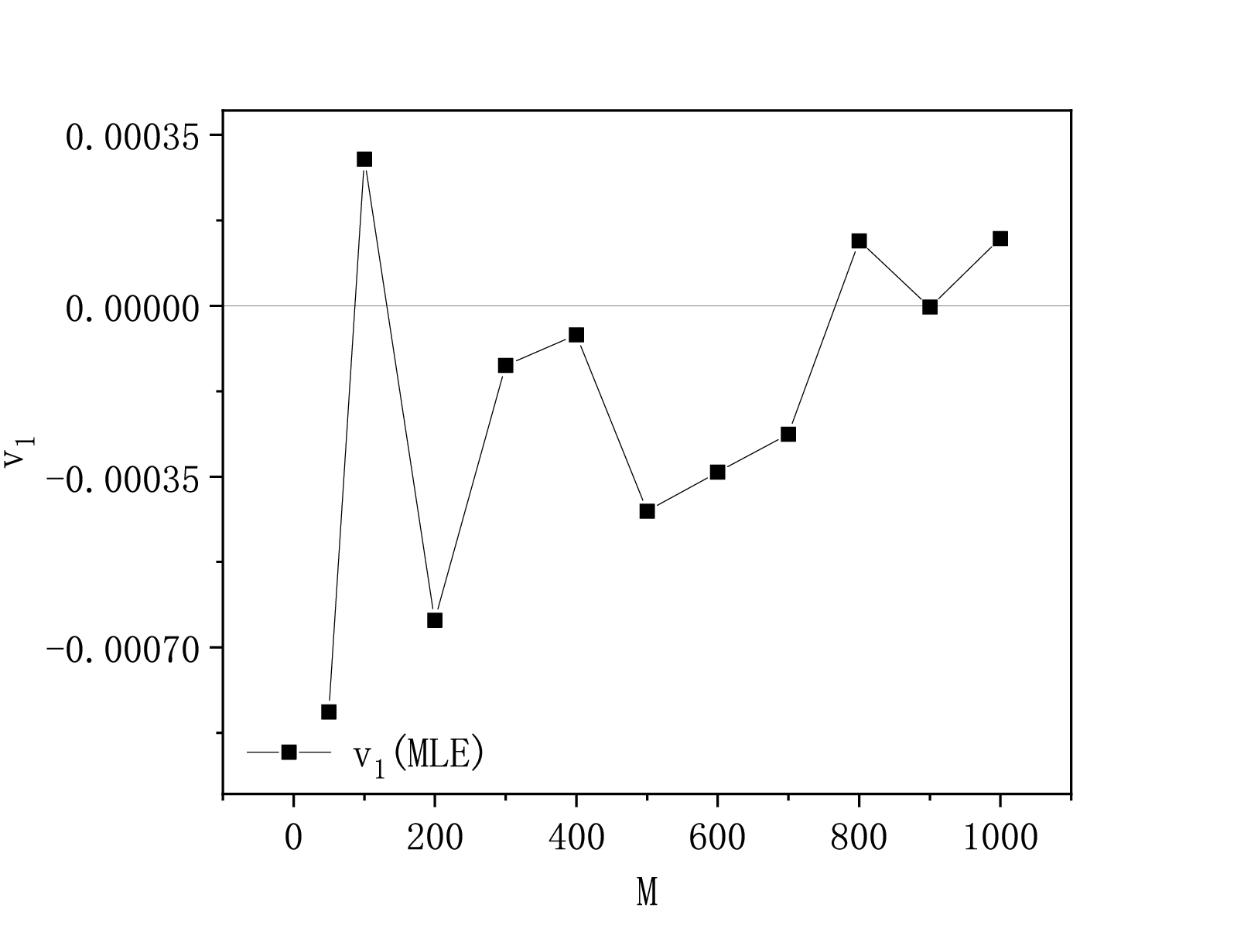}
    \end{minipage}
    \begin{minipage}{0.5\textwidth}
        \centering
        \includegraphics[width=1.0\textwidth, height=0.3\textheight]{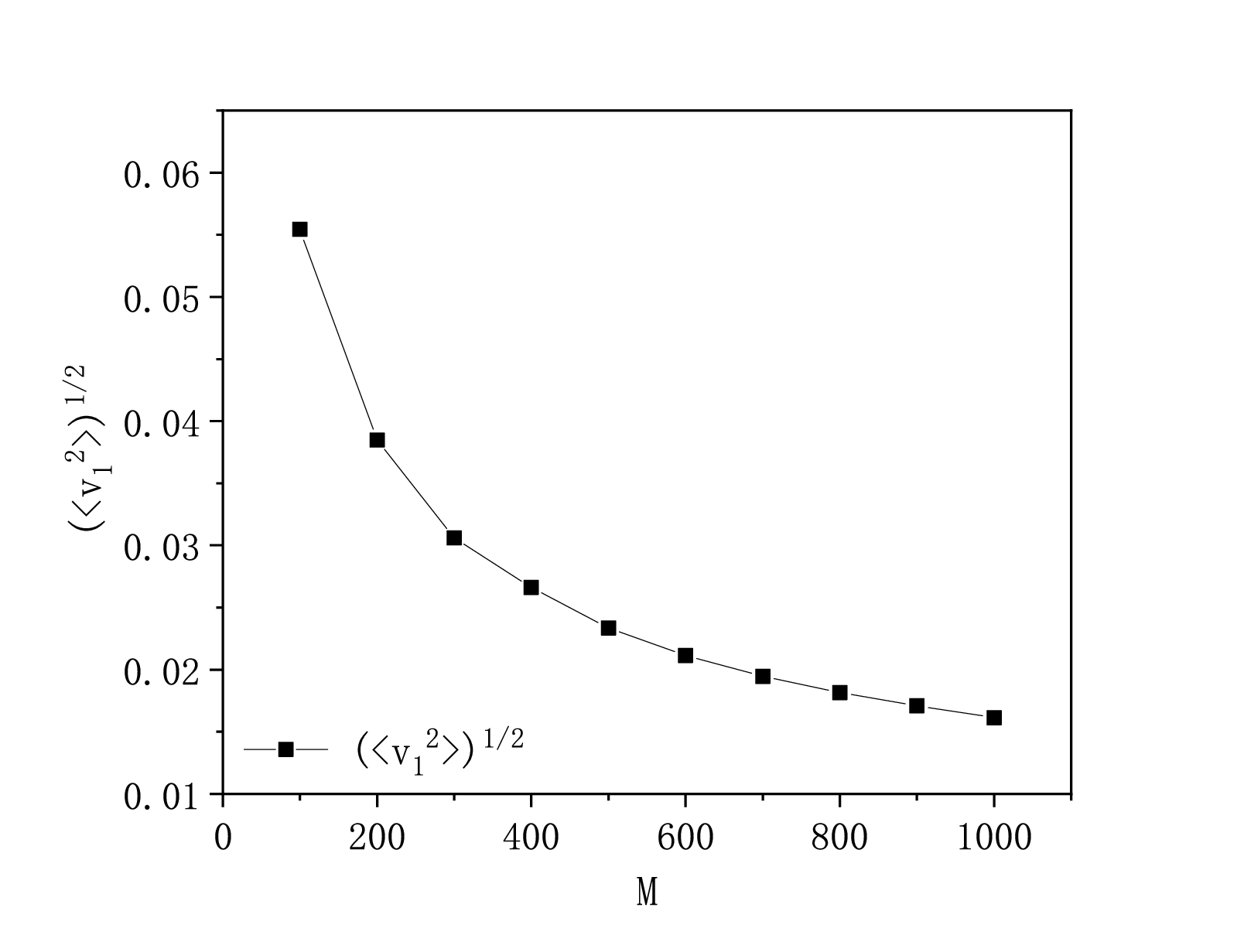}
    \end{minipage}
\renewcommand{\figurename}{Fig.}   
\caption{Correlators and flow harmonics evaluated using events without any background harmonic flow but with independent particle emission under exact global momentum conservation.
The events are generated via the RAMBO algorithm~\cite{jet-ph-05}.
The analysis is performed for 10 000 events across various multiplicities.  
Top: The calculated $c_1\{2\}$ and $c_2\{2\}$ as functions of the multiplicity.  
Bottom: The corresponding event-average MLE estimators $\langle v_1(\mathrm{MLE})\rangle$ and $\sqrt{\langle v_1(\mathrm{MLE})^2\rangle}$ as functions of the multiplicity.}
\label{cmp_scheme2}
\end{figure}

It is noted that the above result differs from those obtained with toy model II. 
As shown in Tabs.~\ref{tabToy2Avg} and~\ref{tabToyratio2}, the deviations from the input values are negative. 
We understand that this is because toy model II proposed above is not identical to the model in Refs.~\cite{hydro-corr-non-flow-01, hydro-corr-non-flow-02}. 
Firstly, the correction to the particle correlation derived in the literature, where non-flow is imposed as a constraint in the phase space integration, depends on the validity of the central limit theorem (or saddle-point approximation), so that the system's total multiplicity cannot be too small. 
In our calculation, however, we have considered smaller clusters with as few as $M_\mathrm{sub}=10$ particles. 
Secondly, unlike Refs.~\cite{hydro-corr-non-flow-01, hydro-corr-non-flow-02}, toy model II is constructed such that not all particles are emitted independently.  
Specifically, compared with an independent-emission Monte Carlo scheme that strictly enforces global momentum conservation, the last few particles generated in toy model II are subject to additional interactions with the rest of the system.  
In view of Eq.~\eqref{vnTrue}, this can induce a nontrivial impact on the flow estimation owing to a nonvanishing true value.  
In this regard, we also performed a flow analysis for events composed of independent particles emitted under exact momentum conservation, implemented via the RAMBO algorithm~\cite{jet-ph-05}.
We evaluate $c_1\{2\}$, $c_2\{2\}$, and $v_1$ using events without any background harmonic flow but with strict global momentum conservation, implemented in the same context as the original study by Borghini, Dinh, and Ollitrault~\cite{hydro-corr-non-flow-01}. 
The analysis is performed for 10,000 events across various multiplicities, and the results are shown in Fig.~\ref{cmp_scheme2}. 
In the top row of Fig.~\ref{cmp_scheme2}, we present the results for $c_1\{2\}$ and $c_2\{2\}$ as functions of the multiplicity $M$. 
It is observed that the results for $c_1\{2\}$~\cite{hydro-corr-non-flow-01} and $c_2\{2\}$~\cite{hydro-corr-non-flow-13} are well reproduced. 
Here, $c_1\{2\}$ is manifestly negative, whereas $c_2\{2\}$ is positive with a significantly smaller magnitude. 
The magnitude of both quantities decreases with increasing multiplicity, consistent with analytical estimates. 
For these same events, we also compute the MLE estimators and plot $\langle v_1(\mathrm{MLE})\rangle$ and $\sqrt{\langle v_1(\mathrm{MLE})^2\rangle}$ as functions of multiplicity, as shown in the bottom row of Fig.~\ref{cmp_scheme2}. 
On the one hand, we see that the magnitude of $v_1(\mathrm{MLE})$ is rather insignificant, namely, $v_1(\mathrm{MLE})\ll \sqrt{|c_1\{2\}|}$, while $\sqrt{c_1\{2\}}$ is not a real number given that $c_1\{2\}<0$. 
This can be understood as follows: the non-flow contribution characterized by $c_1\{2\}$ arises from a nonvanishing $\langle \cos(\phi_1-\phi_2)\rangle \ne 0$ due to momentum conservation, while background collective flow is not included. 
However, under this scenario, although the one-particle distribution is slightly deformed, it remains isotropic, and thus $\langle \cos\phi_1\rangle = 0$. 
On the other hand, the quantity $\sqrt{v_1(\mathrm{MLE})^2}$ is found to decrease with increasing multiplicity, a behavior attributed to the asymptotic property of the MLE, which aligns with the behavior of $\sqrt{c_1\{2\}}$. 
These results support the use of the MLE as a reliable flow estimator.

\subsection{Correction for detector acceptance deficiency}\label{sectionC}

In this subsection, we examine the MLE's ability to correct detector acceptance deficiency in the presence of non-flow.
In reality, a detector's acceptance may vary across its range, potentially introducing appreciable systematic errors in the analysis of anisotropic flow. Several researchers have tackled this challenge in the context of particle correlations and Q-vectors (see, for example, Ref.~\cite{hydro-corr-ph-36}). 
It has been demonstrated that the MLE approach can be modified to accommodate detector inefficiencies~\cite{sph-vn-10}. 
Specifically, the correction of non-uniform detector acceptance is achieved by integrating a weighting scheme directly into the likelihood function.

The extent of nonuniformity in the detector's response can be characterized by the acceptance function $s(\phi)$, which depends on the azimuthal angle and satisfies $s(\phi)\leq 1$, for which the case of perfect detection efficiency corresponds to $s(\phi)=1$. 
For simplicity, any dependence of $s$ on other kinematic variables such as transverse momentum or rapidity will be ignored in the following discussions.

Within the MLE framework, the likelihood function can be compensated for acceptance effects by introducing a weight $w(\phi) = 1/s(\phi)$. 
This leads to a modification to the log-likelihood Eq.~\eqref{eqlogl}
\begin{eqnarray}
{\ell}'(\theta; \phi_1, \phi_2, ..., \phi_M) = \log \mathcal{L}'(\theta; \phi_1, \phi_2, ..., \phi_M)
= \sum_{j=1}^M w(\phi_j) \log f_1(\phi_j; \theta).
\label{eqlikewk}
\end{eqnarray}
Here, the weighting corrects the exponent of the single-particle distribution function, thereby compensating for the suppressed multiplicity in certain regions. 
Calculations using the modified log-likelihood given by Eq.~\eqref{eqlikewk} proceed similarly to the original MLE, where the weighting factor will automatically account for the effects of non-uniform coverage.

To demonstrate the feasibility and utility of this solution, we test it with two examples that feature particular types of detector imperfections similar to those in~\cite {sph-vn-10}.
The first example is a simplified scenario, where the detector's acceptance is given by a piecewise function~\cite{hydro-corr-ph-10}
\begin{eqnarray}
s_1(\phi)= \begin{cases}
0.5\ \quad &\pi/3 <\phi\leq 2\pi/3 \\
1.0\ \quad &\mathrm{ otherwise}
\end{cases} .
\label{eqeffpiecef}
\end{eqnarray}

\begin{figure}[ht]
\begin{tabular}{cc}
\vspace{-26pt}
\begin{minipage}{250pt}
\centerline{\includegraphics[width=1.2\textwidth,height=1.0\textwidth]{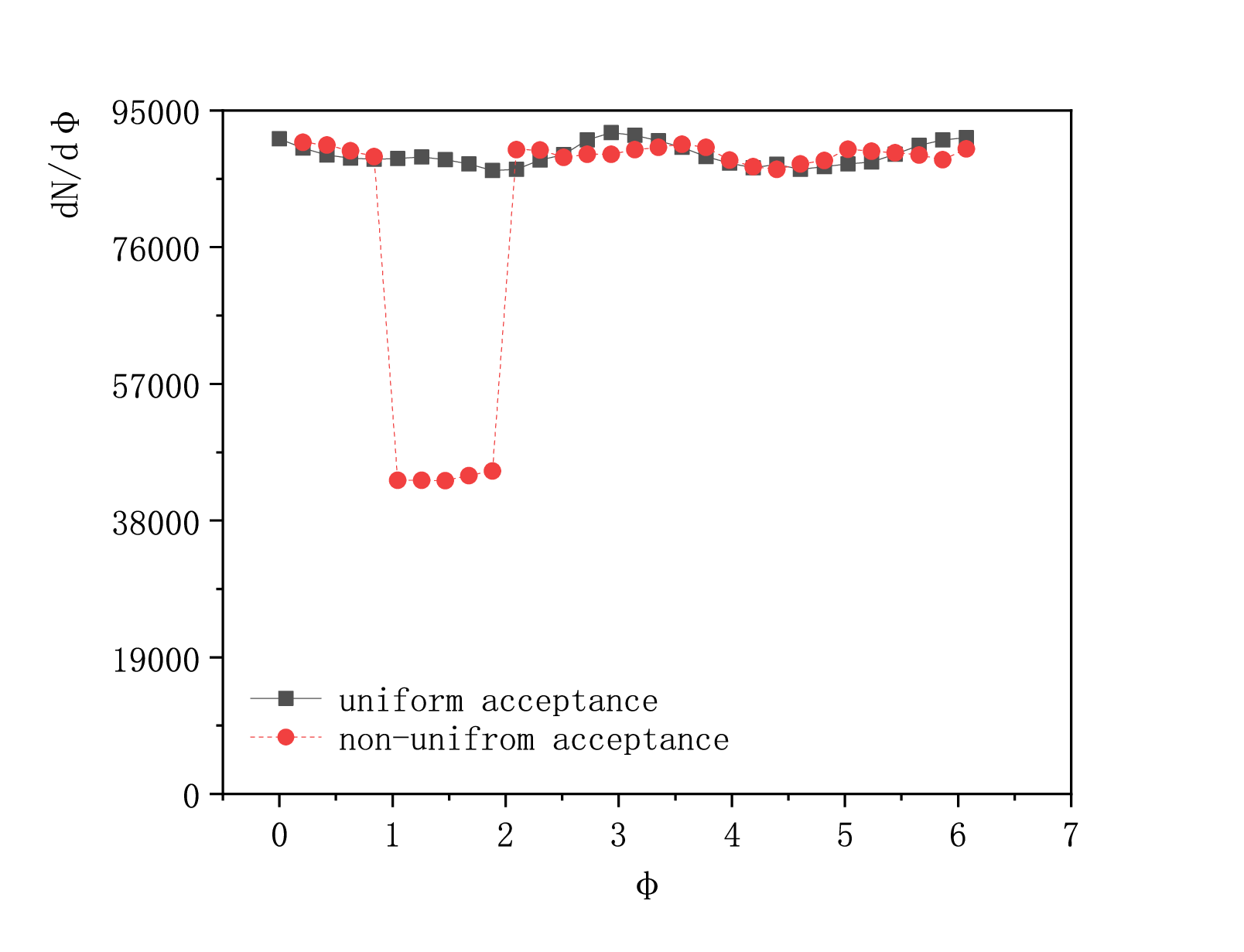}}
\end{minipage}
&
\begin{minipage}{250pt}
\centerline{\includegraphics[width=1.2\textwidth,height=1.0\textwidth]{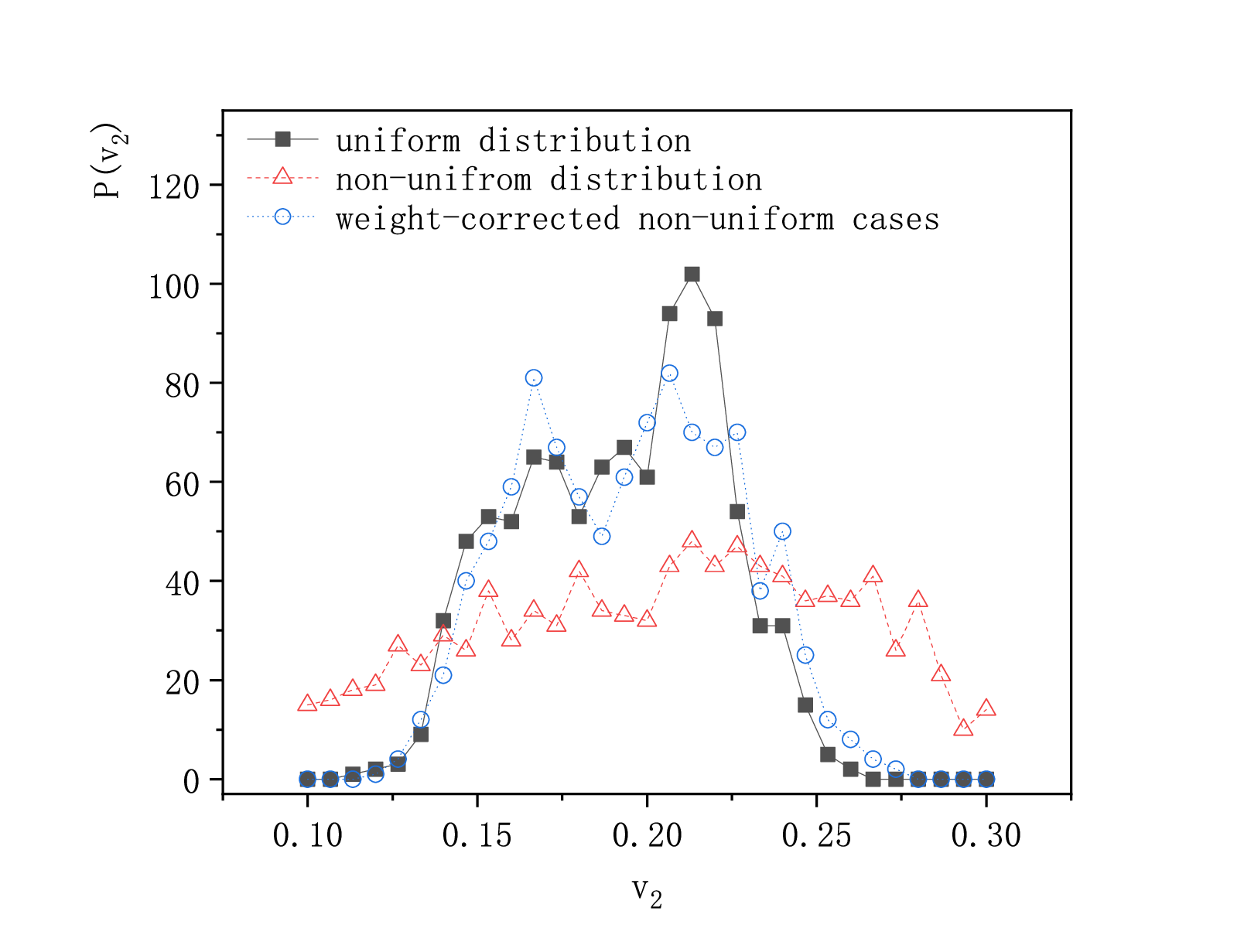}}
\end{minipage}
\\
\vspace{22pt}
\end{tabular}
\vspace{12pt}
\renewcommand{\figurename}{Fig.}
\caption{(Color Online) The results for a simplified detector's acceptance given by a step function Eq.~\eqref{eqeffpiecef}.
For the calculations, a total of 1000 events are generated by toy model I.
Left: The azimuthal particle distribution observed by the detector with a uniform (solid black squares) and non-uniform (solid red circles) acceptance.
Right: The distribution of the estimated $v_2$ on an event-by-event basis using the MLE method, evaluated by considering the correction to the detector's acceptance (empty blue circles).
The results are compared with those obtained without considering the correction (empty red triangles) and with a perfect detector (solid black squares).}
\label{piecefcorr1}
\end{figure}

The results are shown in Fig.~\ref{piecefcorr1}.
In the calculations, we consider 1000 events generated by toy model I with a given multiplicity $M_0=4000$, $v_2=0.2$, $v_3=0.08$, $v_4=0.25$, $P_\mathrm{dcy}=1$, and $\sigma = 15^\circ$.
The MLE approach is employed to estimate the elliptic flow.
In the left panel, we present the distributions of captured particles generated according to the Monte Carlo procedure.
It is observed that the detector's acceptance deficiency leads to an uneven observation in a particular region.
The obtained particle spectra are then utilized by the MLE method to estimate $v_2$.
In the right panel, we show the distributions of the estimated $v_2$ on an event-by-event basis.
It is demonstrated that a non-uniform distribution, if not adequately addressed, can lead to inaccurate estimation. 
The performance can be recuperated appropriately when one adopts the modified likelihood function given by Eq.~\eqref{eqlikewk}.
We note that such a modification can also be generalized to more complicated scenarios, such as the likelihood given in Eq.~\eqref{eqlikelihoodPNew}.

\begin{figure}[ht]
\begin{tabular}{cc}
\vspace{-26pt}
\begin{minipage}{250pt}
\centerline{\includegraphics[width=1.2\textwidth,height=1.0\textwidth]{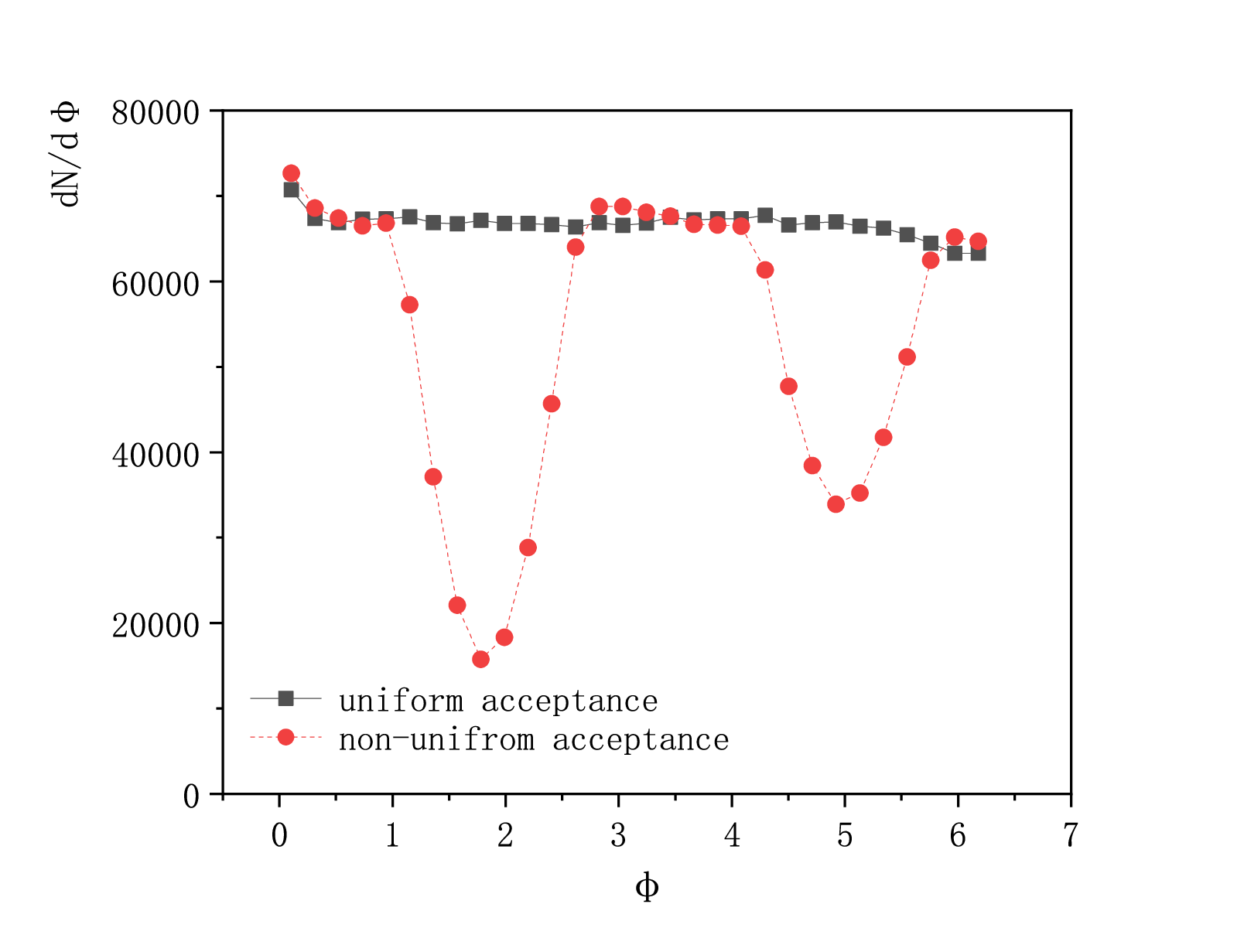}}
\end{minipage}
&
\begin{minipage}{250pt}
\centerline{\includegraphics[width=1.2\textwidth,height=1.0\textwidth]{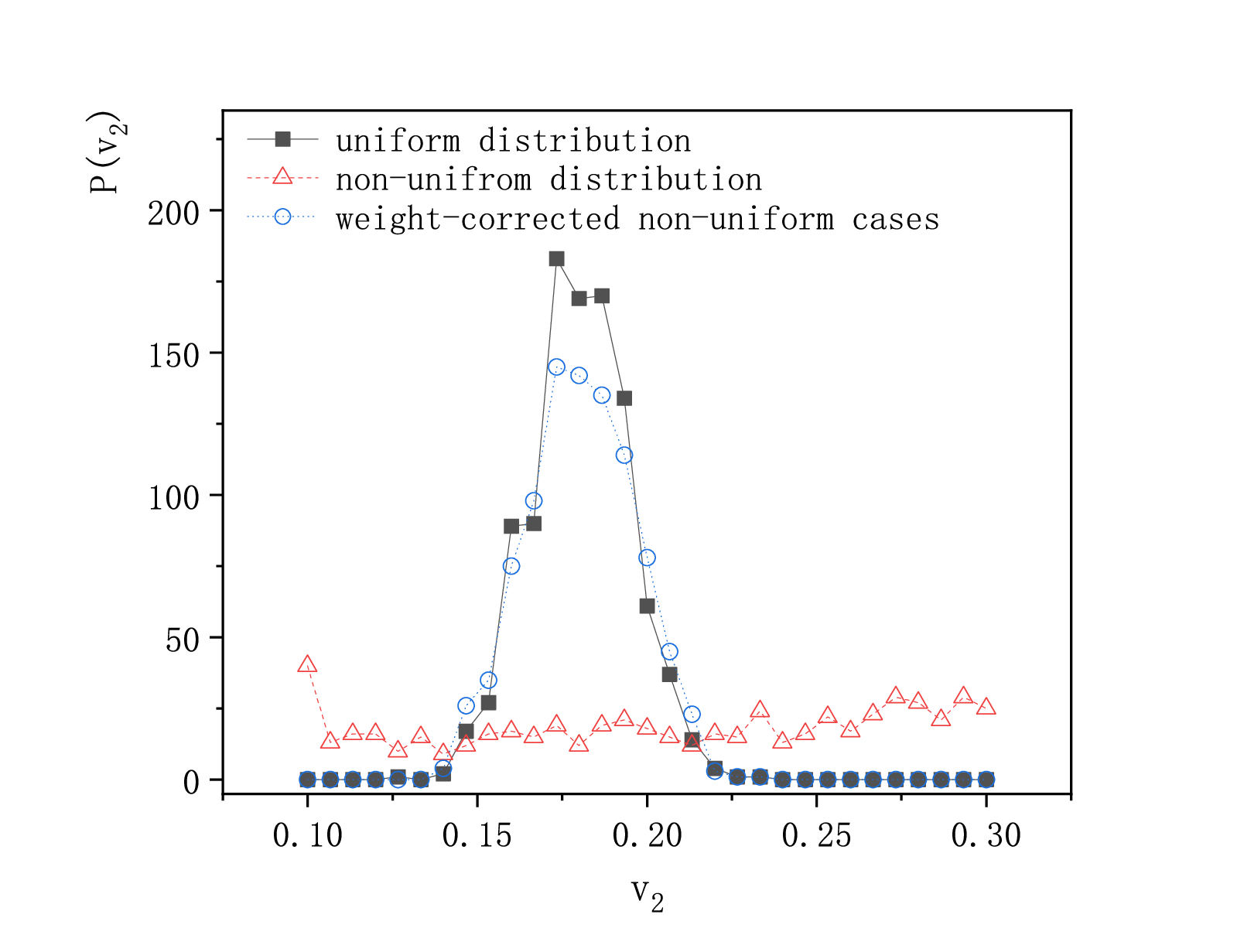}}
\end{minipage}
\\
\vspace{22pt}
\end{tabular}
\vspace{12pt}
\renewcommand{\figurename}{Fig.}
\caption{(Color Online) The same as Fig.~\ref{piecefcorr1} but for the detector's acceptance given by Eq.~\eqref{eqeff2} and the correction scheme is applied to simulated data generated by toy model II.}
\label{piecefcorr2}
\end{figure}

We proceed to elaborate on a more realistic detector's acceptance function given by the following form
\begin{eqnarray}
s_2(\phi)= \begin{cases}
1+\mathrm{exp}(-\phi/7)\sin{2(\phi+0.5)}\ \quad &1.07 <\phi\leq 2.64~\mathrm{or}~4.21 <\phi\leq 5.78\\
1.0\ \quad &\mathrm{ otherwise}
\end{cases}  ,
\label{eqeff2}
\end{eqnarray}
which aims to mimic a detector whose acceptance is suppressed at certain azimuthal angles~\cite{LHC-na49-vn-04}.  
Fig.~\ref{piecefcorr2} shows the numerical results for the data generated by toy model II.
The calculations are carried out using 1000 events with the multiplicity $M\approx 2400$ and $M_\mathrm{sub}\in [3,7]$.
Similarly, the left panel shows the distribution of the detected azimuthal particle spectra.
The right panel displays the distribution of the estimations for $v_2$ for individual events.
While the estimator almost lost its prediction power entirely without considering the correction, the MLE method performs adequately to estimate the elliptic flow, and the probability distribution is consistent with that of a perfect detector.

\section{Concluding remarks}\label{section6}

The present study further generalizes our initial proposal~\cite{sph-vn-10, sph-vn-11} to a more specific subject, namely the non-flow effect.
Our findings suggest that MLE is a meaningful alternative to traditional methods for flow analysis.
In particular, we argue that the two scenarios explored in this study are physically relevant.
On the one hand, regarding particle decay, approximately one-third of the pions measured in heavy-ion collisions originate from specific decay processes, highlighting the relevance of this effect.
The proposed toy model considers an artificial particle pair emission which is correlated to the event plane and thus leads to deviations in the estimated flow different from those in the literature.
On the other hand, although the overall impact of momentum conservation is minor in events with large multiplicity, it may become more significant in small systems or in events where minijets play a prominent role.
However, the strength of MLE lies in its superior accuracy in the limit of large sample sizes, which may not align with the regime where the demand for a new method is greatest.
In this study, we therefore performed simulations for systems with relatively small multiplicities while intentionally amplifying the non-flow effects in the two toy models.
Nevertheless, a key question remains regarding how to effectively apply MLE to systems with even smaller multiplicities, where non-flow effects are physically important.
Also, it is important to note that, except in low-multiplicity events, an effective mechanism for suppressing non-flow effects is to evaluate correlations between particles separated by a pseudorapidity gap, a context which lies beyond the scope of the present study.
We plan to continue investigating these issues in future studies.

\section*{Acknowledgements}

We are thankful for the enlightening discussions with Mike Lisa, who suggested that we explore the non-flow effect resulting from particle decay.
The authors are deeply indebted to Yogiro Hama for his inspiring guidance and unwavering encouragement throughout the years.
We gratefully acknowledge the financial support from Brazilian agencies 
Funda\c{c}\~ao de Amparo \`a Pesquisa do Estado de S\~ao Paulo (FAPESP), 
Funda\c{c}\~ao de Amparo \`a Pesquisa do Estado do Rio de Janeiro (FAPERJ), 
Conselho Nacional de Desenvolvimento Cient\'{\i}fico e Tecnol\'ogico (CNPq), 
and Coordena\c{c}\~ao de Aperfei\c{c}oamento de Pessoal de N\'ivel Superior (CAPES).
A part of this work was developed under the project Institutos Nacionais de Ciências e Tecnologia - Física Nuclear e Aplicações (INCT/FNA) Proc. No. 464898/2014-5.
This research is also supported by the Center for Scientific Computing (NCC/GridUNESP) of São Paulo State University (UNESP).
CY acknowledges the support of the FAPERJ process no. E-26/200.231/2025.

\bibliographystyle{h-physrev}
\bibliography{references_qian}

\begin{thebibliography}{10}

\bibitem{qgp-review-12}
J.-P. Blaizot and E.~Iancu,
\newblock Phys. Rept. {\bf 359}, 355 (2002), arXiv:hep-ph/0101103.

\bibitem{qgp-review-13}
D.~H. Rischke,
\newblock Prog. Part. Nucl. Phys. {\bf 52}, 197 (2004), arXiv:nucl-th/0305030.

\bibitem{RHIC-star-overview-1}
STAR Collaboration, J.~Adams {\em et~al.},
\newblock Nucl.Phys. {\bf A757}, 102 (2005), arXiv:nucl-ex/0501009.

\bibitem{RHIC-brahms-overview-1}
BRAHMS Collaboration, I.~Arsene {\em et~al.},
\newblock Nucl.Phys. {\bf A757}, 1 (2005), arXiv:nucl-ex/0410020.

\bibitem{RHIC-phenix-overview-1}
PHENIX Collaboration, K.~Adcox {\em et~al.},
\newblock Nucl.Phys. {\bf A757}, 184 (2005), arXiv:nucl-ex/0410003.

\bibitem{RHIC-phobos-overview-1}
B.~Back {\em et~al.},
\newblock Nucl.Phys. {\bf A757}, 28 (2005), arXiv:nucl-ex/0410022.

\bibitem{LHC-alice-review-01}
ALICE, K.~Aamodt {\em et~al.},
\newblock JINST {\bf 3}, S08002 (2008).

\bibitem{LHC-atlas-review-01}
ATLAS, G.~Aad {\em et~al.},
\newblock JINST {\bf 3}, S08003 (2008).

\bibitem{LHC-cms-review-01}
CMS, S.~Chatrchyan {\em et~al.},
\newblock JINST {\bf 3}, S08004 (2008).

\bibitem{hydro-review-04}
P.~Romatschke,
\newblock Int. J. Mod. Phys. {\bf E19}, 1 (2010), arXiv:0902.3663.

\bibitem{hydro-review-05}
C.~Gale, S.~Jeon, and B.~Schenke,
\newblock Int. J. Mod. Phys. {\bf A28}, 1340011 (2013), arXiv:1301.5893.

\bibitem{hydro-review-06}
U.~W. Heinz and R.~Snellings,
\newblock Annu. Rev. Nucl. Part. Sci. {\bf 63}, 123 (2013), arXiv:1301.2826.

\bibitem{hydro-review-07}
T.~Hirano, P.~Huovinen, K.~Murase, and Y.~Nara,
\newblock Prog. Part. Nucl. Phys. {\bf 70}, 108 (2013), arXiv:1204.5814.

\bibitem{hydro-review-08}
T.~Kodama, H.~Stocker, and N.~Xu,
\newblock J. Phys. {\bf G41}, 120301 (2014).

\bibitem{hydro-review-09}
R.~Derradi~de Souza, T.~Koide, and T.~Kodama,
\newblock Prog. Part. Nucl. Phys. {\bf 86}, 35 (2016), arXiv:1506.03863.

\bibitem{hydro-review-10}
W.~Florkowski, M.~P. Heller, and M.~Spalinski,
\newblock Rept. Prog. Phys. {\bf 81}, 046001 (2018), arXiv:1707.02282.

\bibitem{RHIC-star-v2-01}
STAR Collaboration, C.~Adler {\em et~al.},
\newblock Phys. Rev. Lett. {\bf 87}, 182301 (2001), arXiv:nucl-ex/0107003.

\bibitem{RHIC-brahms-v2-01}
BRAHMS, I.~Arsene {\em et~al.},
\newblock Nucl. Phys. A {\bf 757}, 1 (2005), arXiv:nucl-ex/0410020.

\bibitem{RHIC-phenix-v2-01}
PHENIX Collaboration, K.~Adcox {\em et~al.},
\newblock Phys. Rev. Lett. {\bf 89}, 212301 (2002), arXiv:nucl-ex/0204005.

\bibitem{RHIC-star-v2-05}
STAR Collaboration, J.~Adams {\em et~al.},
\newblock Phys. Rev. {\bf C72}, 014904 (2005), arXiv:nucl-ex/0409033.

\bibitem{LHC-alice-vn-01}
ALICE Collaboration, K.~Aamodt {\em et~al.},
\newblock Phys. Rev. Lett. {\bf 105}, 252302 (2010), arXiv:1011.3914.

\bibitem{LHC-atlas-vn-01}
ATLAS Collaboration, G.~Aad {\em et~al.},
\newblock Phys. Rev. {\bf C86}, 014907 (2012), arXiv:1203.3087.

\bibitem{LHC-cms-vn-01}
CMS Collaboration, S.~Chatrchyan {\em et~al.},
\newblock Phys. Rev. Lett. {\bf 109}, 022301 (2012), arXiv:1204.1850.

\bibitem{LHC-cms-vn-04}
CMS, V.~Khachatryan {\em et~al.},
\newblock Phys. Lett. {\bf B765}, 193 (2017), arXiv:1606.06198.

\bibitem{LHC-atlas-vn-07}
ATLAS, G.~Aad {\em et~al.},
\newblock Phys. Rev. Lett. {\bf 116}, 172301 (2016), arXiv:1509.04776.

\bibitem{LHC-small-system-review-02}
J.~L. Nagle and W.~A. Zajc,
\newblock Ann. Rev. Nucl. Part. Sci. {\bf 68}, 211 (2018), arXiv:1801.03477.

\bibitem{RHIC-star-v2-10}
STAR, L.~Adamczyk {\em et~al.},
\newblock Phys. Rev. Lett. {\bf 115}, 222301 (2015), arXiv:1505.07812.

\bibitem{hydro-vn-deformed-07}
B.~Schenke, P.~Tribedy, and R.~Venugopalan,
\newblock Phys. Rev. C {\bf 89}, 064908 (2014), arXiv:1403.2232.

\bibitem{hydro-vn-deformed-10}
P.~Carzon, S.~Rao, M.~Luzum, M.~Sievert, and J.~Noronha-Hostler,
\newblock Phys. Rev. C {\bf 102}, 054905 (2020), arXiv:2007.00780.

\bibitem{hydro-vn-deformed-11}
R.~Samanta and P.~Bo\.zek,
\newblock Phys. Rev. C {\bf 107}, 054916 (2023), arXiv:2301.10659.

\bibitem{hydro-vn-deformed-12}
H.~Mascalhusk {\em et~al.},
\newblock Chin. Phys. C {\bf 49}, 054110 (2025), arXiv:2408.06249.

\bibitem{LHC-atlas-vn-20}
ATLAS, G.~Aad {\em et~al.},
\newblock (2025), arXiv:2503.24125.

\bibitem{hydro-v3-02}
D.~Teaney and L.~Yan,
\newblock Phys. Rev. {\bf C83}, 064904 (2011), arXiv:1010.1876.

\bibitem{hydro-vn-33}
D.~Teaney and L.~Yan,
\newblock Phys. Rev. {\bf C86}, 044908 (2012), arXiv:1206.1905.

\bibitem{sph-vn-03}
F.~G. Gardim, F.~Grassi, M.~Luzum, and J.-Y. Ollitrault,
\newblock Phys. Rev. {\bf C85}, 024908 (2012), arXiv:1111.6538.

\bibitem{hydro-vn-34}
H.~Niemi, G.~Denicol, H.~Holopainen, and P.~Huovinen,
\newblock Phys. Rev. {\bf C87}, 054901 (2012), arXiv:1212.1008.

\bibitem{sph-vn-04}
W.-L. Qian {\em et~al.},
\newblock J.Phys.G {\bf G41}, 015103 (2014), arXiv:1305.4673.

\bibitem{sph-vn-06}
F.~G. Gardim, F.~Grassi, P.~Ishida, M.~Luzum, and J.-Y. Ollitrault,
\newblock Phys. Rev. {\bf C100}, 054905 (2019), arXiv:1906.03045.

\bibitem{hydro-vn-45}
J.~Fu,
\newblock Phys. Rev. {\bf C92}, 024904 (2015).

\bibitem{sph-corr-30}
D.~Wen {\em et~al.},
\newblock Eur. Phys. J. {\bf A56}, 222 (2020), arXiv:2004.00528.

\bibitem{sph-corr-33}
S.-F. Shen {\em et~al.},
\newblock Chin. Phys. {\bf 49}, 084104 (2025), arXiv:2502.05737.

\bibitem{event-plane-method-1}
S.~Voloshin and Y.~Zhang,
\newblock Z. Phys. {\bf C70}, 665 (1996), arXiv:hep-ph/9407282.

\bibitem{hydro-vn-08}
J.-Y. Ollitrault,
\newblock Phys. Rev. {\bf D46}, 229 (1992).

\bibitem{hydro-v3-01}
B.~Alver and G.~Roland,
\newblock Phys. Rev. {\bf C81}, 054905 (2010), arXiv:1003.0194.

\bibitem{event-plane-method-2}
A.~M. Poskanzer and S.~A. Voloshin,
\newblock Phys. Rev. {\bf C58}, 1671 (1998), arXiv:nucl-ex/9805001.

\bibitem{hydro-corr-ph-03}
N.~Borghini, P.~M. Dinh, and J.-Y. Ollitrault,
\newblock Phys. Rev. {\bf C63}, 054906 (2001), arXiv:nucl-th/0007063.

\bibitem{hydro-corr-ph-10}
A.~Bilandzic, R.~Snellings, and S.~Voloshin,
\newblock Phys. Rev. {\bf C83}, 044913 (2011), arXiv:1010.0233.

\bibitem{pythia-vn-10}
J.~Jia, M.~Zhou, and A.~Trzupek,
\newblock Phys. Rev. C {\bf 96}, 034906 (2017), arXiv:1701.03830.

\bibitem{hydro-corr-ph-04}
N.~Borghini, P.~M. Dinh, and J.-Y. Ollitrault,
\newblock Phys. Rev. {\bf C64}, 054901 (2001), arXiv:nucl-th/0105040.

\bibitem{hydro-corr-ph-23}
R.~S. Bhalerao, M.~Luzum, and J.-Y. Ollitrault,
\newblock Phys. Rev. {\bf C84}, 034910 (2011), arXiv:1104.4740.

\bibitem{hydro-corr-ph-27}
R.~S. Bhalerao, J.-Y. Ollitrault, and S.~Pal,
\newblock Phys. Rev. {\bf C88}, 024909 (2013), arXiv:1307.0980.

\bibitem{hydro-corr-LY-zeros-01}
R.~S. Bhalerao, N.~Borghini, and J.~Y. Ollitrault,
\newblock Nucl. Phys. A {\bf 727}, 373 (2003), arXiv:nucl-th/0310016.

\bibitem{hydro-corr-LY-zeros-02}
FOPI, N.~Bastid {\em et~al.},
\newblock Phys. Rev. C {\bf 72}, 011901 (2005), arXiv:nucl-ex/0504002.

\bibitem{hydro-corr-LY-zeros-03}
FOPI, N.~Bastid {\em et~al.},
\newblock Phys. Rev. C {\bf 72}, 011901 (2005), arXiv:nucl-ex/0504002.

\bibitem{hydro-corr-ph-36}
A.~Bilandzic, C.~H. Christensen, K.~Gulbrandsen, A.~Hansen, and Y.~Zhou,
\newblock Phys. Rev. {\bf C89}, 064904 (2014), arXiv:1312.3572.

\bibitem{hydro-corr-ph-38}
R.~S. Bhalerao, J.-Y. Ollitrault, and S.~Pal,
\newblock Phys. Lett. {\bf B742}, 94 (2015), arXiv:1411.5160.

\bibitem{hydro-corr-ph-42}
P.~Di~Francesco, M.~Guilbaud, M.~Luzum, and J.-Y. Ollitrault,
\newblock Phys. Rev. {\bf C95}, 044911 (2017), arXiv:1612.05634.

\bibitem{hydro-vn-pca-01}
R.~S. Bhalerao, J.-Y. Ollitrault, S.~Pal, and D.~Teaney,
\newblock Phys. Rev. Lett. {\bf 114}, 152301 (2015), arXiv:1410.7739.

\bibitem{sph-vn-10}
C.~Ye, W.-L. Qian, R.-H. Yue, Y.~Hama, and T.~Kodama,
\newblock Phys. Rev. C {\bf 108}, 024901 (2023), arXiv:2304.00336.

\bibitem{book-statistical-inference-Wasserman}
L.~Wasserman,
\newblock {\em All of Statistics: A Concise Course in Statistical Inference}, 1
  ed. (Springer, 2003).

\bibitem{sph-vn-11}
C.~Ye {\em et~al.},
\newblock Phys. Rev. C {\bf 111}, 034904 (2025), arXiv:2408.14347.

\bibitem{sph-corr-32}
C.~Ye {\em et~al.},
\newblock Universe {\bf 9}, 413 (2023), arXiv:2309.04629.

\bibitem{sph-corr-31}
W.-L. Qian {\em et~al.},
\newblock Universe {\bf 9}, 67 (2023), arXiv:2304.00403.

\bibitem{hydro-corr-non-flow-01}
N.~Borghini, P.~M. Dinh, and J.-Y. Ollitrault,
\newblock Phys. Rev. C {\bf 62}, 034902 (2000), arXiv:nucl-th/0004026.

\bibitem{hydro-corr-non-flow-04}
Q.~Wang and F.~Wang,
\newblock Phys. Rev. C {\bf 81}, 064905 (2010), arXiv:0812.1176.

\bibitem{hydro-corr-ph-09}
J.-Y. Ollitrault, A.~M. Poskanzer, and S.~A. Voloshin,
\newblock Phys. Rev. {\bf C80}, 014904 (2009), arXiv:0904.2315.

\bibitem{hydro-corr-non-flow-review-03}
Y.~Feng and F.~Wang,
\newblock J. Phys. G {\bf 52}, 013001 (2024).

\bibitem{qgp-review-15}
P.~Jacobs and X.~Wang,
\newblock Prog. Part. Nucl. Phys. {\bf 54}, 443–534 (2005).

\bibitem{qgp-review-20}
F.~Wang,
\newblock Prog. Part. Nucl. Phys. {\bf 74}, 35 (2014).

\bibitem{jet-fragmentation-02}
B.~Andersson, G.~Gustafson, and B.~Soderberg,
\newblock Z. Phys. C {\bf 20}, 317 (1983).

\bibitem{hbt-20}
M.~A. Lisa, S.~Pratt, R.~Soltz, and U.~Wiedemann,
\newblock Annu. Rev. Nucl. Part. Sci. {\bf 55}, 357 (2005).

\bibitem{bbc-02}
M.~Asakawa, T.~Csorgo, and M.~Gyulassy,
\newblock Phys. Rev. Lett. {\bf 83}, 4013 (1999), arXiv:nucl-th/9810034.

\bibitem{hydro-corr-non-flow-02}
N.~Borghini, P.~M. Dinh, J.-Y. Ollitrault, A.~M. Poskanzer, and S.~A. Voloshin,
\newblock Phys. Rev. C {\bf 66} (2002).

\bibitem{hydro-corr-non-flow-03}
N.~Borghini,
\newblock Eur. Phys. J. C {\bf 30}, 381 (2003), arXiv:hep-ph/0302139.

\bibitem{hydro-corr-non-flow-07}
Z.~Chajecki and M.~Lisa,
\newblock Phys. Rev. C {\bf 79}, 034908 (2009), arXiv:0807.3569.

\bibitem{hydro-corr-non-flow-08}
N.~Borghini,
\newblock Phys. Rev. C {\bf 75}, 021904 (2007), arXiv:nucl-th/0612093.

\bibitem{hydro-corr-non-flow-13}
A.~Bzdak and G.-L. Ma,
\newblock Phys. Rev. C {\bf 97}, 014903 (2018), arXiv:1710.00653.

\bibitem{hydro-corr-non-flow-14}
A.~Bzdak and G.-L. Ma,
\newblock Phys. Lett. B {\bf 781}, 117 (2018), arXiv:1801.01277.

\bibitem{hydro-corr-non-flow-20}
M.-T. Xie, G.-L. Ma, and A.~Bzdak,
\newblock Phys. Rev. C {\bf 105}, 054904 (2022), arXiv:2204.01038.

\bibitem{hydro-vn-07}
P.~Danielewicz and G.~Odyniec,
\newblock Phys. Lett. B {\bf 157}, 146 (1985), arXiv:2109.05308.

\bibitem{hydro-corr-ph-44}
L.~Nadderd, J.~Milosevic, and F.~Wang,
\newblock Phys. Rev. C {\bf 104}, 034906 (2021), arXiv:2104.00588.

\bibitem{RHIC-star-v2-07}
STAR, C.~Adler {\em et~al.},
\newblock Phys. Rev. C {\bf 66}, 034904 (2002), arXiv:nucl-ex/0206001.

\bibitem{jet-ph-05}
R.~Kleiss, W.~J. Stirling, and S.~D. Ellis,
\newblock Comput. Phys. Commun. {\bf 40}, 359 (1986).

\bibitem{LHC-na49-vn-04}
NA49, C.~Alt {\em et~al.},
\newblock Phys. Rev. C {\bf 68}, 034903 (2003), arXiv:nucl-ex/0303001.

\end{thebibliography}

\end{document}